\documentstyle[prb,twocolumn,aps]{revtex}

\begin{document}
\draft

\title{Properties of lightly doped $t$-$J$ two-leg ladders}

\author{Matthias Troyer$^{a,b,}$\cite{byline}, Hirokazu
Tsunetsugu$^{a,b,c}$, and T.\ M.\ Rice$^{b}$}

\address{
$^{a}$Interdisziplin\"ares Projektzentrum f\"ur Supercomputing, \\
Eidgen\"ossische Technische Hochschule,
CH-8092 Z\"urich, Switzerland \\
$^{b}$Theoretische Physik, Eidgen\"ossische Technische Hochschule,\\
CH-8093 Z\"urich, Switzerland\\
$^{c}$Institute of Applied Physics, University of Tsukuba,
Tsukuba, Ibaraki 305, Japan
}

\date{Received}
\maketitle

\begin{abstract}
We have numerically investigated the doped $t$-$J$ ladder using exact
diagonalization.  We have studied both the limit of strong
inter-chain coupling and isotropic coupling.  The ladder scales to
the Luther-Emery liquid regime in the strong inter-chain coupling
limit.  In this strong coupling limit there is a simple picture of
the excitation spectrum that can be continued to explain the
behavior at isotropic coupling.  At $J=0$ we have indications of a
ferromagnetic ground state.  At a large $J/t$ the ladder is phase
separated into holes and a Heisenberg ladder.  At intermediate
coupling the ground state shows hole pairing with a modified
$d$-wave symmetry.  The excitation spectrum separates into a
limited number of quasiparticles which carry charge $+|e|$ and spin
${1\over 2}$ and a triplet magnon mode.  At half-filling the former
vanish but the latter evolves continuously into the magnon band of
the spin liquid.  At low doping the quasiparticles form a dilute
Fermi gas with a strong attraction but simultaneously the Fermi
wave vector, as would be measured in photoemission, is large.  The
dynamical structure factors are calculated and are found to be very
similar to calculations on 2D clusters.
\end{abstract}


\section{Introduction}
\label{sec:dbl}

The properties of strongly correlated electrons confined to a ladder
(or double chain) and described by $t$-$J$ or Hubbard models have
been the subject of intensive investigation recently.\cite{%
dagotto,noack,tsunetsugu94,rice,sigrist,khveshchenko,tsunetsugu95}
The reason lies in the unusual spin liquid nature of the undoped
parent system.\cite{%
dagotto,hirsch88,strong,gopalan,barnes93,barnes94,white,troyer}
Another reason for especial interest is weakly coupled ladders
compounds like ${\rm SrCu}_2{\rm O}_3 $ and
$({\rm VO})_2{\rm P}_2{\rm O}_7$.\cite{takano,johnston}
Recent measurements of the magnetic susceptibility and the nuclear
spin relaxation rate in these materials show the existence of a
finite spin gap.

The key question in the current study is the evolution of the finite
gap in the spin excitation spectrum upon doping. The spin gap remains
in other spin liquids systems and is a sign of strong superconducting
fluctuations.\cite{ogata_A,imada}

A recent analysis of the $t$-$J$ ladder using a mean-field theory
with Gutzwiller renormalization of the matrix elements to account
for the strong correlations, gave a continuous evolution of the spin
gap with doping.\cite{sigrist}  The short range resonance valence
bond (RVB) state evolves into a superconductor with modified
$d$-wave symmetry within this mean-field approximation.
A tendency towards modified $d$-wave superconductivity was also
found in a bosonization approach\cite{khveshchenko} and in a recent
numerical study of the Hubbard ladder.\cite{noack}

We have investigated $t$-$J$ ladders up to a size of $10 \times 2$
sites using a Lanczos diagonalization method. First results have been
published in Ref.~\onlinecite{tsunetsugu94}.  Here we report in
more detail our results for larger lattices including a detailed
investigation of the excitation spectrum, a discussion of phase
separation and the calculation of the superconducting order
parameter and of the form factor of the Cooper pairs.

We find clear evidence of hole pairing and a modified $d$-wave
RVB state in lightly doped systems in agreement with the mean-field
theory. An interesting difference however is the discontinuous
evolution of the excitation spectrum upon doping.
New ``quasiparticle" excitations appear carrying both charge and
spin.  These excitations are in addition to a band of magnons which
evolve continuously away from the undoped spin liquid.
This separation of the excitation spectrum into bound holon-spinon
quasiparticles and collective magnon excitation contrasts with the
full spin-charge separation found in a Luttinger liquid.

The $t$-$J$ ladder Hamiltonian is
\begin{eqnarray}
\label{eq:ham}
  {\cal H}&=&
   -t\sum_{j,\sigma,a}
       {\cal P}
        \left(
           c_{j,a,\sigma}^\dagger c_{j+1,a,\sigma}
            + {\rm H. c. }
        \right)
       {\cal P}
\nonumber \\ &&
   -t'\sum_{j,\sigma}
       {\cal P}
        \left(
            c_{j,1,\sigma}^\dagger c_{j,2,\sigma}
            + {\rm H. c. }
        \right)
       {\cal P} \\
   &&+J \sum_{j,a}
        \left( {\bf S}_{j,a}\cdot {\bf S}_{j+1,a}
           - \textstyle{1 \over 4} n_{j,a} n_{j+1,a}
        \right)
\nonumber \\ &&
+J' \sum_{j}
        \left( {\bf S}_{j,1} \cdot {\bf S}_{j,2}
          - \textstyle{1 \over 4} n_{j,1} n_{j,2}
        \right), \nonumber
\end{eqnarray}
where $j$ runs over $L$ rungs, and $\sigma$ $(=\uparrow,\downarrow)$
and $a$ $(=1,2)$ are spin and leg indices. The $t$-$J$ ladder is
sketched in Fig.~\ref{fig:geometry}.
The first two terms are the kinetic energies and the $J$ $(J')$ are
exchange couplings along the ladder (rungs).
Unless noted otherwise we set $t'=t$.
The projection operator
${\cal P} \equiv \prod_{i,a}(1-n_{i,a,\uparrow}n_{i,a,\downarrow})$
prohibits double occupancy of a site.  Periodic or antiperiodic
boundary conditions (PBC, APBC) are used along the ladder.
The wave vector ${\bf k}=(k_x,k_y)$ is consequently well defined,
$k_x$ and $k_y$ being the momenta along the ladder and rungs.
The transverse momentum $k_y$ takes only the values $0$ and $\pi$,
corresponding to bonding and antibonding states.

At half filling the $t$-$J$ ladder is equivalent to the Heisenberg
ladder, which was investigated in earlier publications.\cite{%
dagotto,sigrist,gopalan,barnes93,barnes94,white,troyer}
The ground state of the Heisenberg ladder is a short range RVB
state with a spin gap of $\Delta\approx J/2$\cite{%
dagotto,barnes93,white,troyer} at isotropic coupling, $J'=J$.

The strong coupling limit $J'/J\rightarrow\infty$ is a good starting
point to describe the system as there a simple description of the
spectrum is available.\cite{reigrotzki}  In that limit,
each eigenfunction of the total system can be written as a direct
product of one-rung states, which are either spin singlets or one
of the triplets, and the ground state is that with all singlets.
The first excited multiplet consists of the states with one triplet
rung.  A small but finite value of $J$ lifts the degeneracy of
these states. The one-magnon excitations then form a three-fold
spin degenerate band with dispersion
$\epsilon_k=J'+J \cos k_x + {1\over 4} (J^2/J')(3-2\cos 2k_x)$
up to second order in $J$. It has a minimum gap
$\Delta = J'-J+{1\over 2} {J^2\over J'}$
at $k_x=\pi$.\cite{reigrotzki}
The momentum perpendicular to the chains is $k_y=\pi$. The higher
excited states form a continuum of excited states and its minimum
is at ${\bf k}=(0,0)$ with energies slightly larger than twice
the gap $2\Delta$.  With increasing $J$ the collective excitation
branch crosses into the continuum, but the qualitative description
is still valid.

In this paper we study the effects of doping holes into such a
ladder.  Although the isotropic case, $J'/J=1$, is of most interest,
we also study the limit $J' \gg J, t$, which can be easily
understood.  In this limit the problem reduces to a system of
weakly coupled rungs. The properties can be continuously followed
down to the isotropic point $J=J'$.

This paper is organized as follows. In Sec.~\ref{sec:ferro} we
briefly discuss the occurrence of ferromagnetism in the ladder
doped with one or two holes at $J=0$ and discuss the relationship
with the occurrence of ferromagnetism in two dimensions.
Next in Sec.~\ref{sec:pair} we discuss the pairing of holes doped
into the ladder and the occurrence of phase separation.
To understand the excitation spectra we start from the
single hole case in Sec. \ref{sec:onehole} and go on to the two-hole
case in Sec.~\ref{sec:twohole}. Section~\ref{sec:paircorr} discusses
long range correlations, in particular the interesting question of
the symmetry of the pairs and the mapping to a Luther-Emery liquid.
The single-particle excitations are discussed in
Sec.~\ref{sec:chargexc}.  Over all we find a remarkable similarity
between the ladder and 2D clusters.

\section{Ferromagnetism for $J=0$}
\label{sec:ferro}

The $t$-model ($t$-$J$ model with $J=0$) is equivalent to the
infinite-$U$ Hubbard model. In single chains the ground state of the
$t$-model is degenerate in the spin degrees of freedom. In two
dimensions on the other hand the ground state of the $t$-model doped
with one hole is ferromagnetic.\cite{nagaoka}  This is called the
Nagaoka effect.

The extension of the proof by Nagaoka to finite hole doping in the
thermodynamic limit proved to be difficult. Actually the ground state
of the two-dimensional (2D) square-lattice $t$-model doped with
{\it two} holes is {\it not} ferromagnetic.\cite{doucot}  For finite
densities in the thermodynamic limit there are contradicting
results. Variational estimates for the $U=\infty$ Hubbard model
indicate that the fully polarized ferromagnetic state is stable until
a critical doping $\delta_{cr}=0.29$.\cite{linden}  High
temperature series expansions by Putikka {\it et al.} on the other
hand show evidence that the fully polarized ferromagnetic ground
state does not survive at any finite doping. Instead they find
evidence for a partially polarized ferromagnetic state at low
hole doping.  A fully polarized ferromagnetic state at finite
doping was found only for $J<0$.\cite{putikka}

In this context it is of interest to study the occurrence of
ferromagnetism in the ladder models. While the proof by Nagaoka
\cite{nagaoka} cannot be applied to the one-dimensional chain
it is valid for the ladder. The proof relies on the existence of
closed loops on the lattice. Such loops exist in 2D
lattices and on ladders, but cannot be formed on single chains. The
ground state of the ladder doped with one hole is thus
ferromagnetic.

We have numerically studied the $t$-ladder with
$L=2,3,\ldots,10$ rungs, doped with two holes. In
Fig.~\ref{fig:FM_engy} we show the ground state energies of the
ladders for both PBC's and APBC's.
We find that the ground state is always ferromagnetic for
APBC and an even number of rungs and for PBC and an odd number of
rungs. For the other boundary conditions the ground state is a spin
singlet.

An important point is that the ferromagnetic state always has the
lower energy for a ladder with at least four rungs. The singlet state
is very close in energy and deserves a more detailed investigation.
In Fig.~\ref{fig:FM_corr} we plot the real-space spin correlations
$\langle S^z(0)S^z(r)\rangle$ of the lowest singlet state of the
$L=10$ ladder. These spin correlations show that the singlet state
actually consists of two ferromagnetic domains with opposite
magnetization.

The results show clear evidence for a ferromagnetic ground state of
the $t$-ladder ($L \ge 4$) doped with two holes. In the thermodynamic
limit however two holes is not a finite density.  Extrapolations of
our small-cluster results at finite doping to the thermodynamic limit
$L \rightarrow \infty$ are hard to obtain. But one may speculate that
the existence of a ferromagnetic ground state of the $t$-ladder
with 2 holes and $L\ge4$ could indicate a ferromagnetic state for
dopings $\delta < \delta_{\rm cr} \approx0.25$.  Similar results
were obtained by Hirsch and M\"uller-Hartmann.\cite{hirsch_phd}

\section{Hole Pairing and Phase Separation}
\label{sec:pair}

\subsection{Hole Pairing}

In this section we will discuss the pairing of holes doped into the
$t$-$J$ ladder and the occurrence of phase separation at large values
of $J/t$. We will start from the simple limit $J'\gg J,t$. In this
limit the undoped ladder consists of weakly coupled rungs, as is
sketched in Fig.~\ref{fig:strong_limit}(a).

In this limit two holes doped into the ladder will go onto the same
rung in order to minimize the number of broken singlet bonds.  This
state is graphically shown in Fig.~\ref{fig:strong_limit}(c).
In order to study the occurrence of hole pairing at smaller values
of $J'$ and down to the isotropic point $J=J'$ we calculate the
binding energy and the hole-hole correlation function. We find
that even at isotropic coupling the holes still form a bound pair,
although the pair is more spread out there.

The binding energy $E_B$ is defined as
\begin{equation}
   E_B \equiv 2E_{\rm G.S.}(2L-1) - E_{\rm G.S.}(2L )
            - E_{\rm G.S.}(2L-2),
\end{equation}
where $E_{\rm G.S.}(N)$ is the ground state energy for $N$ electrons,
the boundary conditions are chosen between PBC and APBC to give the
lowest energy.

In the large $J'$ region the binding energy can easily be estimated.
A single hole doped into a Heisenberg ladder breaks one bond
with energy loss $J'$, but can gain kinetic energy $-t$ along the
ladder (see the next section for details) and $-t'$ along the rung.
It follows that
$E_{\rm G.S.}(2L-1)\approx E_{\rm G.S.}(2L)+J'-t-t'$.
Two holes on the same rung also break one bond, but the
kinetic energy of such a bound pair is much smaller, of order
$-4t^2/J'$, as will be calculated later. Thus we estimate
$E_{\rm G.S.}(2L-2)\approx E_{\rm G.S.}(2L ) +J'$,
and a binding energy:
\begin{equation}
   E_B \approx J'-2t-2t'\qquad \mbox{for}\;J'\gg J,t,t'\;.
\end{equation}

Figure \ref{fig:gaps} shows $E_B$ as a
function of $J'$.  It remains positive and thus shows binding down to
the isotropic value, $J/t=J'/t=0.3$.  The same holds for a larger
$J/t=0.5$.

Additional evidence for pairing is provided by the hole-hole
correlation functions
\begin{equation}
   \bigl\langle  n_{\rm h}(0) n_{\rm h}(r) \bigr\rangle \equiv
   \bigl\langle  (1-n_{i,a}) (1-n_{i+r,a'})
   \bigr\rangle,
\end{equation}
measured on the same leg $a=a'$ and on different legs $a\ne a'$
in the ground state.  They are plotted in Fig.~\ref{fig:hole_corr}
for $J/t=0.3$ and $J'/J=1$ and $10$.  For $J'/J \gg 1$ the two
holes are predominantly on the same rung and the correlation
function shows a clear exponential decay. At the isotropic point
the pair is more extended. The maximum of the correlation function
is now at a distance 1 along the legs and on different legs,
but it again decays at large distances.  We can calculate the
size of the hole pair by fitting the inter-chain correlations
to an exponential form
$\langle n_{\rm h}(0) n_{\rm h}(r) \rangle
  \sim e^{-r/\xi} + e^{-(L-r)/\xi}$
for the two largest distances, $L/2$ and $L/2-1$.
The inset of Fig.~\ref{fig:hole_corr} shows the size $\xi$
of the hole pair as a function of the inter-chain coupling $J'/t$.
The pair is very tightly bound for $J'\gg J$. At the isotropic
point the pair is still bound, with a diameter of about two lattice
spacings.  Note the oscillation of the radius with respect to $L$.
The size seems to converge to a value in between the $L=8$ and the
$L=10$ result at the isotropic point.

\subsection{Effective boson model for the large $J'$ limit}

We may say that the system belongs to the Luther-Emery universality
class of 1D correlated systems,\cite{luther_A} in the sense that
the spin excitations acquire a finite gap while the charge
excitations remain gapless.  In the limit of large $J'$, however,
the picture that tightly bound hole pairs are moving in a
background of singlet rungs is more appropriate than weak coupling
approaches like $g$-ology.  Considering these hole pairs as hard
core bosons, we can determine the long-range correlations by a
mapping to an effective boson model.

The pair hopping matrix element to second order in
perturbation theory is
\begin{eqnarray}
   t^* = { 2t^2 \over J'-{4t'^2 / J'}}.
\end{eqnarray}
There is a weak attraction $V^*$ between two hole pairs on
neighboring rungs, which again to second order takes the form
\begin{eqnarray}
   V^* \equiv -{ J \over 2 } - {3J^2 \over 8J'} +
               {4t^2 \over J'-{4t'^2 / J'}},
\end{eqnarray}
where the first, attractive, term comes from the charge part of the
$J$-term in the Hamiltonian. As $t^* , V^* \ll J'$ we can map the
low-energy part of the $t$-$J$ ladder onto an effective hard-core
boson model on a chain with nearest neighbor interaction:
\begin{equation}
   H^{*} =
   -t^* \sum_i \bigl( B_i^{\dag} B_{i+1} + \mbox{H. c.} \bigr)
   +V^* \sum_i N_i N_{i+1},
\end{equation}
where the hard-core boson creation operator $B_i^{\dag}$ creates a
hole pair at the rung $i$ and $N_i \equiv B_i^{\dag}B_i$ is its
number operator.  There is a hard-core repulsion since only one
hole pair can be created on any given rung.

Our effective boson model is equivalent to the XXZ-model in a
magnetic field, which has been solved exactly by a bosonization
approach and conformal
field theory.\cite{haldane}  For $V^* < -2|t^*|$ the system is phase
separated. This is the case for $J' > J'_{\rm PS}$, where
\begin{equation}
  J'_{\rm PS} =  { 16 \, t^2 \over J } - {J \over 2} +
     {\rm O} \left( {J^3 \over t^2} \right),
\end{equation}
again to second order perturbation theory.  For physically reasonable
values of $J/t$ phase separation occurs only at very large values of
$J'$: $J'_{\rm PS}/t=53.2$ for $J/t=0.3$ and $J'_{\rm PS}/t=31.8$ for
$J/t=0.5$.  Note that the dominant attractive part of the interaction
comes from the charge part $-{1\over4} J n_{j,a} n_{j+1,a}$ of the
$J$-term.

Next we will discuss the region where the system is not yet
phase separated but $J'$ is still large ($J,t \ll J' < J'_{PS}$).
There we can determine the dominant correlations from the
effective boson model.  The correlation exponents have been
calculated indirectly by Bethe ansatz.\cite{haldane}  Both the
charge density wave correlations and the superconducting
correlations show a power-law decay at large distances:
\begin{mathletters}
\begin{eqnarray}
   \langle N_r N_0\rangle &&\sim
   {\rm const.} \times r^{-2} +
   {\rm const.} \times \cos \bigl( 2k_F r \bigr) \, r^{-K_\rho},\\
   \langle B_r^{\dag} B_0\rangle &&\sim r^{-1/K_\rho}.
\end{eqnarray}
\end{mathletters}
The superconducting correlations $\langle B_r^{\dag}B_0\rangle$ are
dominant if $K_\rho>1$. This is the case for most of the phase
diagram, except for the phase separation regime at $V^*<-2t^*$. At
quarter filling $\rho=1/2$ and for $V^*>2t^*$ the system is in the
Ising-limit and shows a long range charge density wave ground
state. At fillings close to that line and for large $V^*>2t^*$ there
is a small region where $K_\rho<1$.\cite{haldane}

In our effective model we have $V^*<0$ and there
$K_\rho>2$.\cite{haldane}  We are thus always in the region of
dominant superconducting correlations. Even neglecting the
attractive charge part of the $J$-term we are still in the
superconducting regime where $K_\rho>1$.

In the limit of large $J'$ the equivalence of the $t$-$J$ ladder with
a Luther-Emery liquid can clearly be seen.  Going to isotropic
coupling the spin gap remains finite and the only low-lying
excitation is the collective charge mode, as we will show in the
following sections.  Thus also at isotropic coupling the $t$-$J$
ladder is still a Luther-Emery liquid.  In Sec. \ref{sec:map},
we will develop another approach which relates the long-range
correlations to thermodynamic quantities for more general $J$'s,
based on a bosonization of density fluctuations.

\subsection{Phase separation}

Finally we study the occurrence of phase separation at isotropic
coupling $J=J'$. We estimate the onset of phase separation by
determining the coupling $J$ at which the compressibility $\kappa$
diverges. The compressibility per site can be calculated as usual
\begin{equation}
   \kappa^{-1}=\rho^2 \
   { \partial^2 \epsilon (\rho)
     \over
     \partial \rho^2 }\;,
\label{eq:kappa}
\end{equation}
where $\epsilon(\rho)$ is the energy density per site of the ladder
with a particle density per site $\rho = N/(2L)$.

In a finite system usually the discrete version
\begin{equation}
   \kappa^{-1}={N^2\over 2L}
   \left[
     { E(N+2;L)+E(N-2;L)-2E(N;L) \over 4}
   \right]
\label{eq:kappaf}
\end{equation}
is used, where $E(N;L)$ is the ground state energy of the finite
system with $N$ particles on the ladder with $L$ rungs (Volume $2L$).
At small hole doping however this procedure may not be reliable due
to finite size effects caused by frustration on small lattices.
To see this let us consider the $L=8$ ladder doped with zero,
two or four holes.  In the undoped case there are 8 spins on each
leg of the ladder.  Two holes doped into the ladder will
predominantly go onto different legs and there will be seven spins
per leg. Thus the antiferromagnetic configuration on the legs will
be frustrated.  For four holes there will be six holes on each leg
and the system is again not frustrated. Conversely on an $L=9$
ladder the undoped ladder and the ladder doped with four holes will
be frustrated, while the ladder doped with two holes will be
non-frustrated.

We have thus used a different formula to calculate the
compressibility at small hole doping. We calculate the ground state
energies for an $L=8$ ladder doped with $N_{\rm h}=0$ and $4$ holes
and for an $L=9$ ladder doped with $2$ and $6$ holes.  In all these
cases the ladder
is not frustrated. Then we estimate the compressibility from these
energies using finite differences similar to the above
Eq. (\ref{eq:kappaf}). In the thermodynamic limit
$L \rightarrow \infty$ both formulas give the same result,
as the frustration appears only on small lattices.

While the finite size effects are quite small at low electron
densities they are much larger at small hole dopings due to
frustration mentioned before. The estimated errors on the phase
separation line may thus be much larger there, about $\pm 0.2t$.

A comparison with the results obtained with open boundary
conditions (OBC) confirms our results. Only at small doping
the OBC results are not reliable since there the holes are trapped
on the ends of the chain.

Figure \ref{fig:phasesep} shows the phase separation line for the
$t$-$J$ ladder for $J=J'$, in the $J$-$\rho$ plane. Note that,
opposite to the single chain case,\cite{ogata_B} the onset of phase
separation at small hole doping is at lower values of $J/t$ than at
small electron concentrations. This resembles the behavior in two
dimensions,\cite{putikka} although the precise position of the
phase separation line in two dimensions has not yet been established.

\section{Properties of a single hole}
\label{sec:onehole}

In the previous section we have discussed the ground state of
the ladder doped with two holes. In order to understand the low
energy excitations of the ladder it is useful to study the one-hole
problem first.

As mentioned above the limit $J'\gg J,t$ is a good starting point to
explore the $t$-$J$ ladder.  There are nine different states,
depicted in Fig.~\ref{fig:nine_basis}.  A single electron goes
either into the bonding or antibonding orbital
\begin{equation}
   b_{i,\sigma}^{\dag} = {1\over\sqrt{2}}
   \left(
      c_{i,1,\sigma}^{\dag} + c_{i,2,\sigma}^{\dag}
   \right),
   a_{i,\sigma}^{\dag} = {1\over\sqrt{2}}
   \left(
      c_{i,1,\sigma}^{\dag} - c_{i,2,\sigma}^{\dag}
   \right),
\end{equation}
with energy $ \mp t'$, respectively. Two electrons on the rung are
either in the singlet state with energy $-J'$ or in one of the three
triplet states with energy~$0$.
The singlet state expressed in bonding and antibonding orbitals is
\begin{equation}
   {1 \over \sqrt{2}}
   \left(
      c^{\dag}_{i,1,\uparrow} c^{\dag}_{i,2,\downarrow}
    - c^{\dag}_{i,1,\downarrow}c^{\dag}_{i,2,\uparrow}
   \right) =
   {1 \over \sqrt{2}}
   \left(
      b_{i,\uparrow}^{\dag}b_{i,\downarrow}^{\dag}
    - a_{i,\uparrow}^{\dag}a_{i,\downarrow}^{\dag}
   \right)
\end{equation}
Similarly the three triplets can be expressed as combinations of one
bonding and one antibonding electron:
\begin{equation}
   a^{\dag}_\uparrow b^{\dag}_{\uparrow},\;
   {1\over\sqrt{2}}
   \left(
      a^{\dag}_\uparrow b^{\dag}_{\downarrow}
     +a^{\dag}_\downarrow b^{\dag}_{\uparrow}
   \right),\;
   a^{\dag}_\downarrow b^{\dag}_{\downarrow}
\end{equation}

Figures \ref{fig:disp_1h}(a) and (b) show the one-hole spectra for
$L=8$ for large inter-chain coupling $J'/J=10$, calculated by
exact diagonalization for $J/t=0.3$, $J'/t=3$ and $J/t=0.5$, $J'/t=5$
respectively.

A hole on a single rung can be either in the bonding or the
antibonding orbital. One hole doped into the half filled ladder
will thus be either in a bonding or antibonding state, depending on
the parity symmetry of the total ladder
[see Fig.~\ref{fig:strong_limit}(b)].  This hole can
propagate along the ladder with a hopping matrix element
$\tilde{t}=+t/2$ in first order perturbation theory.  Thus the low
energy states are two bands of holes in the bonding and anti-bonding
orbitals.  They are split by the energy difference $2t'$ of the
bonding and antibonding states.  These two bands can clearly be seen
in the spectra [Figs.~\ref{fig:disp_1h}(a) and (b)].  The minimum of
the bands is at $k_x=\pi$, since the hopping matrix element for
holes $\tilde{t}>0$.  The bandwidth of both bands is $4\tilde{t}=2t$
in the limit $J'\gg J,t$.  At finite $J$ the bandwidth is reduced
due to hybridization with the higher excited states.

Decreasing $J'$ to the isotropic points $J=J'$ changes the dispersion
of these bands [see Figs.~\ref{fig:disp_1h}(c) and (d)].  At low
energies we can still see the bands of holes in the bonding and
antibonding orbitals. These bands evolve continuously from the large
$J'$ limit. The minima of the energy bands are not at $k_x=0$ or
$k_x=\pi$, but at a large momentum
${\bf k}_{F}^{B}\approx\left(\pm{3\over5}\pi,0\right)$ for the
bonding and
${\bf k}_{F}^{A}\approx\left(\pm{2\over5}\pi,\pi\right)$ for the
antibonding band.  We can fit the low-lying hole bands to
a dispersion of the form
\begin{eqnarray}
   E(k_x) = &&E_0+\Delta E+\alpha_1\cos k_x \nonumber\\
   &&\ \ \ + \alpha_2 \cos 2k_x + \alpha_3 \cos 3k_x,
\label{eq:fitspec}
\end{eqnarray}
corresponding to nearest neighbor ($\alpha_1$), next-nearest neighbor
($\alpha_2$) and third-nearest neighbor ($\alpha_3$) hopping.  $E_0$
is the ground state energy of the undoped ladder and $\Delta E$ the
shift in energy of the center of the band upon doping.  In
Fig.~\ref{fig:disp_fit} we show the bands and the excellent fit.
The parameters are shown in Table \ref{tab:fitspec}.

The changes in the hole dispersion with decreasing $J'$ are
summarized as follows:

(i) The center of the bands shifts downwards by $\Delta E < 0 $,
compared to the undoped ladder. The energy gain for one hole in the
case of $J=0$ would be just the kinetic energy $-t'$.  When $J > 0$
we lose magnetic energy by introducing the hole.  The energy gain
is therefore smaller at larger $J/t$, as we can also see from the
fit parameters.

(ii) The hole bands are narrowed compared to the large $J'$ limit.
In that limit the bandwidth of the hole bands was $2t$.
This bandwidth is renormalized by the stronger polarization effects
at isotropic coupling, and it is now of the same order as the
magnetic energy $J$, instead of the kinetic energy $2t$.

(iii) The dispersion changes as longer range hopping processes
($\alpha_2,\alpha_3)$ are introduced with decreasing $J'$, and the
minima move away from $k_x = \pi$.  The minima of both bands are
very close in energy, again in contrast to the strong coupling
region where they are split by $2t'$.  In Sec. \ref{sec:chargexc}
we will identify the minima $k_{F}^{B}$ and $k_{F}^{A}$ with the
Fermi points of the bonding and antibonding quasiparticle bands.

Another interesting question is the behavior of the free spin that is
left over after one hole has been doped into the ladder.
In the $t$-$J$ chain the spin and charge excitations are carried by
different soliton excitations which are far apart in space from each
other.  This is a typical feature of spin-charge separation and such
a system is called a Luttinger liquid.  In a Fermi liquid on the
other hand they are bound and the excitations are described by
quasiparticles carrying both charge and spin.

We have calculated the hole-spin correlations to answer the question
if spin-charge separation occurs in the ladder.  The real space
correlations
\begin{equation}
   \langle n_{h,a} (j+r) S^{z}_{a'} (j) \rangle ,
\end{equation}
are shown in Fig.~\ref{fig:hole-spin}.  This correlation function is
nonzero for the ground state in the subspace of
$S^z_{\rm tot} ={1 \over 2}$ since there remains one spin unpaired.

The result shows that the hole is tightly bound to the remaining
free spin.  At strong inter-chain coupling $J'\gg J,t$ it is again
predominantly on the same rung. At isotropic coupling the spin-hole
bound state is more extended. These spin-hole bound states thus
carry both charge and spin.  In this sense they are similar to the
quasiparticles in a Fermi liquid.  This is in contrast to the
spin-charge separation in the single chain. We will therefore call
the single holes bound to the free spin ``quasi-particles'',
although the system has a spin gap.

\section{Excitation spectra of the ladder with two holes}
\label{sec:twohole}

\subsection{Excitation Spectra}

The ground state of the ladder doped with two holes is, as discussed
above, a bound state of the two holes. This bound pair coherently
propagates along the ladder, giving rise to the lowest-lying band.
When $J'\gg J$, this band, spin-singlet charge excitations, is
clearly seen in the numerical results
as shown in Fig.~\ref{fig:disp_2h_strong}.

The higher energy excitations are again understood simply
in the large $J'$ limit.  An essential difference from the
lowest-lying singlet band
is that two holes are now separate rather than forming a bound
pair.  Being separate, they can gain a larger kinetic energy,
but only in return for a even larger cost of exchange energy
$\sim J'$ as one more singlet rung is
broken. Thus there are continua of scattering states of the two holes
(``quasiparticles'') at higher energies.  Since the residual
interactions between the two quasiparticles are weak,
the energy is almost degenerate between the $S=0$ and
$S=1$ spin subspaces.
On the finite lattice we naturally do not
see a continuum of scattering states, but only several discrete
bands. These bands, and the fact that the energies of the triplet and
singlet are nearly degenerate (up to boundary effects) can be seen in
the spectra.

There are various combinations of the two quasiparticle bands
in the two quasiparticle continuum of states.
The lowest are scattering states of two bonding
quasiparticles, with $k_y=0$. Higher states are scattering states of
one bonding and one anti bonding quasiparticle.  Having the same
transverse momentum, $k_y=\pi$
these states hybridize with the one-magnon excitations in the
spin background, resulting in a ``bound state'' below the
two-particle continuum.  This is nearly dispersionless and clearly
seen in the spectrum.  The continua of two antibonding
quasiparticles is much higher in energy and not included in the
figure.

At higher energies there are spin excitations in the spin background.
They are not described by the quasiparticles and
will be discussed in more detail in the next section.

The results in the isotropic case, $J'/J=1$, is shown in
Fig.~\ref{fig:disp_2h_iso}.
The energy spectrum is more complicated but the above description
still holds qualitatively.

The ground state is still the bound hole pair. It moves coherently
along the ladder, yielding a gapless band of singlet charge
excitations.  The band has a linear dispersion around
${\bf k}=(0,0)$ compared with quadratic in the large $J'$ case.
An important point is that despite its complicated dispersion the
low energy part is well separate from the other excitations.
Therefore also at isotropic coupling the only low-energy excitation
is the collective charge excitations, and we may identify the
isotropic $t$-$J$ ladder as Luther-Emery liquid.  We will discuss
the essential role of these charge fluctuations concerning
superconductivity in Sec.~\ref{sec:map}.

In addition to the gapless band of charge fluctuations, there are
various local minima at higher energies.  However, they can be
explained by taking account of the nonmonotonic dispersion of the
one-hole spectra shown in Fig.~\ref{fig:disp_1h}, and our
quasiparticle picture still holds.  More specifically, there are
four local minima in the single-hole spectra at
${\bf k} \approx ( \pm {3 \over 5} \pi,0)$ and
${\bf k} \approx ( \pm {2 \over 5} \pi,\pi)$, which have nearly the
same energy.  Thus to construct low-energy two-quasiparticle
excitations there are many possible combinations of different
minima of one-particle states as discussed below.  This is the
origin of many local minima in the two-particle spectra and this is
confirmed by the fact that the dependence on the boundary condition
is consistent with this picture.

The ground state of the two-hole spectrum can be constructed from
holes near $\pm k_F$ in the single-hole spectra. The other minima
can be explained similarly. The minimum in the PBC spectrum at
$k_x ={4\over5}\pi$ can be identified with the $2k_F$ excitation,
where a particle moves from one Fermi point to the opposite one.
($2 \times { \pm 3 \over 5} \pi \equiv {\pm 4 \over 5} \pi$).
When using APBC we do not have the $k$-value of $\pm{3\over5}\pi$
in the single-hole spectrum on an $L=10$ ladder.  The closest
$k_x$-points are $\pm{5\over10}\pi$ and $\pm {7 \over 10} \pi$,
leading to minima at $\pm \pi$ and $\pm {3 \over 5} \pi$ in the
two-hole spectra. Another feature that can be explained from the
single hole spectra is the minimum at ${\bf k}=(\pi,\pi)$
(odd parity, $k_x=\pi$). Using PBC this state is obtained with one
hole with ${\bf k}=({3\over5}\pi,0)$ and one with
${\bf k}=({2\over5}\pi,\pi)$, leading to the minimum at $(\pi,\pi)$.

Using APBC's we can combine one hole at
${\bf k}=({3 \over 10} \pi,\pi)$ with one at either
${\bf k}=({5 \over 10} \pi,\pi)$ or
${\bf k}=({7 \over 10} \pi,\pi)$.  As mentioned above these two
states are higher in energy than the minimum at
$\pm {3 \over 5} \pi$ and very similar in energy.  Therefore we
expect two states at
${\bf k}=({4 \over 5} \pi,\pi)$ and ${\bf k}=(\pi,\pi)$ which
are similar in energy but at a higher energy than the corresponding
states with PBC. This is exactly what we observe.
The odd parity states near $k_x =0$ can be constructed similarly.

The qualitative features of the three-hole excitation spectrum can
again be explained similarly.

\subsection{Spin Excitations}
\label{sec:spinexc}

One of the most interesting properties of the $t$-$J$ ladder is that
there are two distinct types of spin excitation.  Although it is
most easily seen in the large $J'$ limit, the qualitative
distinction remains down to the isotropic point.

The first type is the collective magnon excitations inherited from
the undoped spin ladder.  One of the electron-filled rungs is now
excited to a spin triplet.  This local excitation is what we call
``magnon'' and it propagates coherently along the ladder, leading to
an energy dispersion with respect to $k_x$.  For a detailed
investigation, we have examined the two-hole spectra in more detail
and have calculated the spin-spin and spin-hole correlations of the
low-lying triplet states.   We find that the magnon excitations of
the Heisenberg ladder evolve continuously upon doping.  However
there discontinuously appears a new kind of spin-triplet excitation
at lower energies, which is not present in the undoped ladder.

The lowest excitation is a different type for which quasiparticles
play an essential role.  Therefore, the spin gap, defined as the
excitation energy to the lowest triplet, is a discontinuous
function of the hole doping at $\delta =0$.  This new type of
excitations consists of breaking a pair of holes into two separate
quasiparticles, each carrying charge $+|e|$ and spin
$1/2$.\cite{K_takano}  When the two quasiparticles are both in the
bonding orbital, their lowest energy, at ${\bf k}=(0,0)$, is lower
than the lowest magnon excitation.  The additional energy gain is
easily understood, since the two separate holes have a larger
kinetic energy of the order of $t$ while the gain of the magnon
kinetic energy is the order of $t^2 /J'$.  As was shown in
Fig.~\ref{fig:gaps}, down to the isotropic point the lowest
two-quasiparticle excitation is lower in energy than the lowest
magnon excitation.

The above picture is confirmed by comparing the correlation function
between the two different states.  The dynamical spin structure
factor gives another confirmation.

Figure \ref{fig:corrs} shows the equal-time correlations of the
two holes,
$\langle n_{\rm h} ({\bf r}) n_{\rm h} ({\bf 0}) \rangle$,
and of spin and hole,
$\langle S^z ({\bf r}) n_{\rm h} ({\bf 0}) \rangle$,
calculated for these two types of spin-triplet excitations.
The latter quantity is nonzero since the states with $S^z =1$
are used in the calculation.  When $J'/J=10$, the two holes
are separate in space in the lowest state, while tightly
bound in the other state.  The position of magnetic excitation is,
on the other hand, close to the hole position in the lowest state,
while they are far apart from each other in the other state.
These two behaviors are what is predicted by our picture explained
above, and despite modification in small detail they are
qualitatively consistent even at the isotropic point.

Our numerical results confirm that the lowest triplet state is
the quasiparticle excitation where the bound hole pair breaks up.
The two holes repel each other and the hole-hole correlations have
the maximum at the largest distance $L/2$.  Each of the holes is
bound to a spin-$1/2$, as can be seen from the hole-spin correlation
function. This state is sketched in Fig.~\ref{fig:strong_limit}(d).

As a typical magnon excitation, we show the correlations for the
state at ${\bf k}=(\pi,\pi)$ that has the main spectral weight in
the dynamical spin structure factor which will be discussed soon.
The hole-hole correlations are similar to the ground state.
The hole-spin correlations show that the triplet carrying spin
current is far away from the hole pair.  This state is shown in
Fig.~\ref{fig:strong_limit}(e).  Mean field theory\cite{sigrist}
predicts only this magnon excitation, which evolves continuously
{}from the Heisenberg ladder.

In neutron scattering experiments the relevant quantity is the
dynamical spin structure factor
\begin{eqnarray}
   {\cal S}({\bf q},\omega_0) \equiv  \sum_{n}
   \Bigl| \langle n|S_{\bf k}^{z}|\mbox{G.S.}\rangle
   \Bigr|^2
   \, \delta(E_n-E_{\rm G.S.}-\omega_0),
\end{eqnarray}
where $|n\rangle$ is the complete set of eigenstates with energy
$E_n$, $|\mbox{G.S.}\rangle$ is the ground state with energy
$E_{\rm G.S.}$ and
\begin{equation}
   S_{\bf q}^{z} \equiv {1\over\sqrt{2L}}\sum_{\bf r}
   e^{i {\bf q} \cdot {\bf r}} S_{\bf r}^{z} .
\end{equation}

Figure \ref{fig:spin_0h} shows ${\cal S}({\bf q},\omega)$ calculated
for the Heisenberg ladder using the Lanczos diagonalization combined
with the continued fraction method.\cite{heine}  It can clearly be
seen that the dominant contributions arise from the collective
excitations near ${\bf q}=(\pi,\pi)$ and there is very little weight
in the continuum of spin excitations at higher energy.

In the doped case the two types of spin excitations have different
contributions to the dynamical spin structure factor. It can be seen
in Fig.~\ref{fig:spin_2h} that the continuum of spin excitations
move towards lower energies.  Most of the weight is in the magnon
excitations, consistent with the mean-field theory,\cite{sigrist}
and there is very little weight in the lowest triplet excitation
consisting of the two separate quasiparticles.

As in the large $J'$ limit the lowest triplet excitation with
$q_y=\pi$ is a bound state of a spin triplet and the hole pair.
At ${\bf q}=(\pi,\pi)$ this state has no spectral weight,
while most of the weight is in the second excited state,
which has the triplet separated from the bound pair.
This is a finite size effect of the two-hole system.
The reason is that at $q_y=0$ or $\pi$ we have an additional
symmetry, reflection invariance in the ladder direction.
The parity under these reflections is different for the ground
state and the triplet-hole pair ground state, leading to the
vanishing weight.

A significant difference between the two types of spin excitations
is that the largest number of ``quasiparticle" excitations is
limited by the number of holes. We need at least two holes to
create such an excitation.  The number of possible excitations
is thus proportional to the hole doping $\delta$.  On the other
hand, the magnon excitations can be excited at any rung where
there are no holes.  The number of these excitations is thus
proportional to $1-\delta$ instead, much larger for a small doping
$\delta$.  Therefore with decreasing temperature the susceptibility
will show a large exponential drop at temperatures of the order
of the gap of the undoped system ($T\sim0.5J$), followed by a
small drop at temperatures around the spin gap of the doped system.

To summarize we can describe the spin excitations of the $t$-$J$
ladders by a simple picture: quasiparticles moving in a spin
liquid background.  In the ground state the quasiparticles are
paired.  In the excitation spectrum two types of excitations
can be distinguished.  The first corresponds to the breaking
of a pair of quasiparticles.  This excitation has the lowest
energy, but its number is limited by the number of holes. Of more
importance for measurements of the susceptibility or inelastic
neutron scattering experiments is the second type, which are magnon
excitations in the spin liquid background.  They evolve continuously
{}from the undoped Heisenberg ladder.
Although the gap for this type of excitation is larger it is more
important since we can excite more of these excitations.
Also the weight in the dynamical spin structure factor is larger.

\subsection{Charge Excitations}

Similarly to the calculation of the dynamical spin structure factor
we calculate the dynamical charge structure factor defined by
\begin{equation}
   {\cal N}({\bf q},\omega_0)\equiv \sum_{n}
   \Bigl| \langle n|\rho_{\bf q} |\mbox{G.S.} \rangle
   \Bigr| ^2
   \delta( E_n - E_{\rm G.S.} - \omega_0),
\end{equation}
where $\rho_{\bf q}$ is the Fourier transform of the local density
fluctuation around the average density $\rho$.

The result of our calculations is shown in
Fig.~\ref{fig:charge_2h}.  The main contribution arises from the
coherent motion of the hole pairs. This leads to large peaks
at low energies near ${\bf q}=(0,0)$.  The rest of the weight is
distributed incoherently over a wide region
at rather high energies of order $4t$.  It arises from interactions
of a single hole with the surrounding spin background.  This is
similar to results obtained for two dimensions.\cite{eder}

Recently Tohyama et al. have calculated ${\cal S}({\bf q},\omega)$
and ${\cal N}({\bf q},\omega)$ for single chains and 2D
clusters.\cite{tohyama}  In the single chain they find that, as
expected from Luttinger liquid theory, the charge and spin
excitations are decoupled in the low energy region.  The dynamical
charge structure factor is very similar to that of spinless
fermions, consisting of large peaks at the energies expected
{}from the cosine band of the spinless fermions.

In two dimensions they find different behavior.  In the spin
structure factor nearly all of the weight is in a few sharp peaks
at low energies.  In the charge structure on the other hand the
main weight is at relatively large energies, of the order of
several $t$, and it strongly broadened.  This is indicative of
strong spin-charge interactions.  Only at some special
$\bf q$-points are there peaks at low energies.  The dynamical
charge structure factor, ${\cal N}({\bf q},\omega)$ for
the ladder shown in Fig.~\ref{fig:charge_2h} differs a lot from the
single chain but resembles the 2D cluster.  At large $q_x$, we see
the peaks at large energies $\sim 2t \mbox{-} 4t$.
At small $q_x$ in the $q_y=0$ sector, ${\cal N}({\bf q},\omega )$
for the ladder is dominated by the collective mode of the hole
pairs, and its behavior differs substantially from spinless
fermions.  Another similarity to the 2D system is found in the
dynamical spin structure ${\cal S}({\bf q},\omega)$ where the
main weight is in peaks with energies $\sim J$
(see Fig.~\ref{fig:spin_2h}).  We conclude from this comparison
that the 2D clusters and the ladders are closely related -
a fact which points towards a ``d-wave'' paired state for the
2D clusters.  A more difficult question is to what extent
this behavior of the finite 2D cluster is a consequence of the
strong tendency of finite clusters to favor singlet ground states
and to what extent it is representative of an infinite plane.

\section{Long Range Correlations}
\label{sec:paircorr}

\subsection{Superconducting Correlations}

A highly intriguing and much debated subject of the high-$T_c$
superconductors is the internal symmetry of the order parameter.
In this section we study the internal structure of the pairs in the
doped ladders.  For the $t$-$J$ model usually only nearest-neighbor
pairs have been considered except for a few cases. It is reasonable
to assume that they are the dominant pair correlations.
However more quantitatively we have chosen the optimal form for
the pairs.

Using the Lanczos algorithm we calculate the pairing correlation
functions for different pairings.  Let us introduce the operator
creating a singlet pair of electrons on sites ${\bf r}$ and
${\bf r+d}$,
\begin{equation}
  P_{\bf r,d}^{\dag}={1 \over \sqrt{2}}
  \left(
    c_{{\bf r},\uparrow}^{\dag} c_{{\bf r+d},\downarrow}^{\dag}
   -c_{{\bf r},\downarrow}^{\dag} c_{{\bf r+d},\uparrow}^{\dag}
  \right),
\end{equation}
Using this definition of the pair operator we can calculate the
superconducting order parameter
\begin{equation}
   \chi_{\bf d} =
   \Bigl\langle
      \mbox{G.S., $N_{\rm h}$$-$2 holes}
   \Big|
      {1\over 2L} \sum_{\bf r} P_{\bf r,d}^{\dag}
   \Big|
      \mbox{G.S., $N_{\rm h}$ holes}
   \Bigr\rangle
\end{equation}
and its Fourier transform
\begin{eqnarray}
   \chi_{\bf k}&&=
   \sum_{\bf k} \chi_{\bf d} \, e^{i {\bf k} \cdot {\bf d} }
   \nonumber \\
   &&=
   \bigl\langle
      \mbox{G.S., $N_{\rm h}$$-$2 holes}
   \big|
      P_{{\bf k},-{\bf k}}^{\dag}
   \big|
      \mbox{G.S., $N_{\rm h}$ holes}
   \bigr\rangle
\end{eqnarray}
where $P_{{\bf k},-{\bf k}}$ is the Fourier transform of the real
space pair operator with zero total momentum.

Figure \ref{fig:pair_k} shows the superconducting order parameter
for the $L=8$ ladder with PBC's for several values of $J/t$.
The most obvious properties are that the sign is opposite for pairs
with $k_y=0$ and $k_y=\pi$ and that the absolute values are very
small near $(0,0)$ and $(\pi,\pi)$. This is similar to $d$-wave
pairing in a fully 2D system.  The absolute value is largest near
wave vectors which we have identified with the Fermi points of the
quasiparticles
${\bf k}=(3 \pi /4,0) \approx {\bf k}_{F}^{B}$ and
${\bf k}=(\pi /2,\pi)\approx {\bf k}_{F}^{A}$.  At large $J/t$
the order parameter is very similar to the simple
$\cos k_x-\cos k_y$ structure of nearest neighbor $d$-wave pairs.

Consider now pairs with a form factor $F$:
\begin{equation}
   P_{\bf r}^\dagger
   = \sum_{\bf d} F_{\bf d}^* \, P_{{\bf r},{\bf d}}^\dagger,
\end{equation}
where $F$ is normalized
\begin{equation}
   \sum_{\bf d} |F_{\bf d}|^2 \equiv 1.
\end{equation}
A simple calculation shows that the superconducting
``order parameter''
\begin{equation}
   \chi_F = \sum_{\bf d} F_{\bf d}^* \, \chi_{\bf d}
\end{equation}
is maximized by the pair with the form factor
\begin{equation}
   F_{\bf d} =
   { \chi_{\bf d}
     \over
     \Bigl[
       {\displaystyle \sum_{\bf r}|\chi_{\bf r}|^2 }
     \Bigr] ^{1/2}
   },
\label{eq:formfactor}
\end{equation}
which is proportional to the order parameter $\chi_{\bf d}$.
This optimal order parameter then is
\begin{equation}
   \chi = \Bigl[ \sum_{\bf d}|\chi_{\bf d}|^2 \Bigr] ^{1/2}.
\label{eq:optimal}
\end{equation}

In Fig.~\ref{fig:OP} we show this maximal order parameter for a
$L=8$ ladder as a function of $J/t$. A clear increase as a function
of $J/t$ can be seen, similar to single chains and 2D planes where
the superconducting correlations are also enhanced at larger $J$
values.  We note that the expectation value of the optimal order
parameter is larger for four doped holes than for two holes.
This is a consequence of the strong correlation effect, as any
pairing order parameter or fluctuations should vanish as
$\delta \rightarrow 0$.  The mean-field theory including the strong
correlation effects predicts a $\delta$-linear dependence of the
order parameter as $\delta \rightarrow 0$:\cite{sigrist}
i.e., the superfluid density is proportional to the hole doping.
Our results show that the value for four doped holes is less than
twice the value for two holes.  This may be because that in the
four hole case the doping, $\delta=0.25$, is already large and out
of the region of $\delta$-linear dependence.  We need further study
using larger lattices to clarify this point more quantitatively.

\subsection{Mapping to a Luther-Emery Liquid}
\label{sec:map}

In Secs.~\ref{sec:pair}-\ref{sec:twohole}, our numerical results
have shown that all the spin excitations cost a finite energy and
the only gapless excitations are charge fluctuations,
i.e., coherent propagation of bound hole pairs along the ladder
direction.  We may therefore say that the system is a Luther-Emery
liquid\cite{luther_A} from large $J'$ limit down to the isotropic
coupling.  This observation indicates that the low-energy and
long-wavelength properties of lightly doped $t$-$J$ ladders would
be entirely described by an effective continuum Hamiltonian
in terms of charge degrees of freedom and this model correctly
predicts long-range asymptotic behavior of correlation functions.
The effective model is actually the bosonic Gaussian model proposed
by Efetov and Larkin,\cite{efetov}
\begin{eqnarray}
  \label{Efetov_Larkin}
   {\cal H}_{\rm EL}
   &&= {\pi \over 2} \int dx
   \left\{
     { [ n_B (x) - \overline{n}_B ]^2   \over  \pi K_{\rm B}  }
     +  \pi K_{\rm B} v_s^2
     \Bigl[
        { \nabla \theta_B (x)  \over \pi }
     \Bigr] ^2
   \right\},
\end{eqnarray}
where $n_B (x)$ represents the density of bound hole pairs
at the rung $x$ with the average value
$\overline{n}_B \equiv N_B / L = \delta$, and
$\theta_B (x)$ is its conjugate phase obeying the canonical
commutation relation,
$[n_B (x) , \theta_B (x')] = i \delta (x-x')$.
As shown by the fact that the $q=0$ mode of the first term
in Eq.\ (\ref{Efetov_Larkin}) denotes the change in the ground
state energy associated with the number of bound hole pairs,
$ \Delta E \propto (1/2K_B) (\Delta N_B)^2 $,
the parameter $K_{\rm B}$ is given by the compressibility of the
hole pairs, which will be shown afterwards.  The second term,
on the other hand, describes the energy change associated with
current, related with the sound velocity $v_s$.  Instead of a
direct calculation of the dispersion relation, the sound velocity
can also be obtained numerically by applying an Aharanov-Bohm flux
penetrating in the center of the ladder with PBC's, since the flux
induces a finite current along the chain direction.  Once the two
parameters, $v_s$ and $K_{\rm B}$, are determined in this way,
it is straightforward to calculate the power-law exponents of
correlation functions as follows.

The Efetov-Larkin Hamiltonian is actually identical to the
single-component (i.e., spinless) Luttinger model diagonalized by
bosonization, which was studied by Mattis and Lieb,\cite{mattis}
Luther and Peschel,\cite{luther_B} and Haldane\cite{haldane}
in details.
The Hamiltonian (\ref{Efetov_Larkin}) is immediately solved by
rewriting each Fourier component with a boson operator,
\begin{eqnarray}
  \label{LE_bosons}
   n_{B,k} \equiv
   {\displaystyle \sqrt{{ |k| \over 2 \pi}} }
   \bigl( b_{k} + b_{-k}^\dagger \bigr) , \ \
   \theta_{B,k} \equiv i
   {\displaystyle \sqrt{\pi \over 2 |k|} }
   \bigl( b_{k}^\dagger - b_{-k}  \bigr) ,
\end{eqnarray}
then the result is
\begin{eqnarray}
  \label{Ham_diag}
   {\cal H}_{\rm EL} = \sum_{|k|}
    v_s |k| b_{k}^\dagger b_{k}
    +{\pi \over 2L}
        \Bigl[ v_N (N_{\rm B}-N_{\rm B}^0)^2
              +v_J J_{\rm B}^2
        \Bigr] .
\end{eqnarray}
Here $N_B$ is the total number of bound hole pairs and
$v_N= (\pi K_{\rm B})^{-1}$ is the charge velocity and
$v_J = \pi K_{\rm B} v_s^2$ is the current velocity associated with
the number of $2k_F^{\rm SF}$ excitations, $J_{\rm B}$, obeying the
universal relation, $v_N v_J = v_s^2$.
Here $k_F^{\rm SF} = \pi \delta = \pi (1-\rho)$ is the Fermi wave
number of spinless fermions transmuted from the bosonic hole pair
operators by the Jordan-Wigner transformation.  The last two terms
in Eq.\ (\ref{Ham_diag}) describe non-bosonic excitations
accompanied with the change in the quantum numbers, $N_{\rm B}$
and $J_{\rm B}$.  Their importance was first pointed out by Haldane
in his Luttinger liquid concept.\cite{haldane}

The propagator of bound hole pair, i.e. pairing correlation, is then
obtained as in the calculation of the Debye-Waller factor,
\begin{eqnarray}
  \label{pair_corr}
   G(r) &&=
   \overline{n}_B \, \langle e^{i [ \theta_B (r) - \theta_B (0) ]}
              \rangle
   = \overline{n}_B \, e^{ - \langle [ \theta_B (r) - \theta_B (0) ]^2
                      \rangle /2 } \nonumber\\
   &&\propto r^{-1/K_{\rho}},
\end{eqnarray}
where the exponent is
\begin{equation}
  \label{exponent}
  K_{\rho} = 2 \pi K_{\rm B} v_s =
  {\pi} {\rho}^2 \, v_s \, \kappa,
\end{equation}
where $\rho$ is the electron density per site and $\kappa$ is the
electron compressibility.  The relation for the ladder,
$K_B = \kappa \rho^2 /2$, is used to obtain the second equality.
This result differs from the one for the chain, which has the
numerical factor $\pi /2$ instead.  It is important to emphasize
that this result is
the Luttinger liquid parameter, $K_\rho = 2 (v_J / v_N)^{1/2}$, for
the single-component boson systems.

Density-density correlations are characterized by a power-law decay
at the wave number $2k_F^{\rm SF}$.  Since $k_F^{\rm SF}$ is twice
the average of the Fermi wave number of the original electron
bonding and antibonding bands, $k_F = {\pi \over 2} \rho$, the wave
number $2k_F^{\rm SF}$ is actually $4k_F$ in the original picture.
The exponent of the $2k_F^{\rm SF}$=$4k_F$ CDW fluctuations is
calculated using a density-phase duality as was done in the
original paper.\cite{efetov}
The density operator has a short-range $2k_F^{\rm SF}$ CDW order,
but since the value of $2k_F^{\rm SF}$ is locally determined by the
density at each position, it is fluctuating around the average value
$2\pi \delta$ associated with the density fluctuations.
We therefore may write as
\begin{mathletters}
  \label{density}
\begin{eqnarray}
   n_B (x) &&= \overline{n}_B \cos
   \bigl[ 2k_F^{\rm SF} x + 2 \theta_J (x) \bigr], \\
   {\partial \theta_J \over \partial x}
   &&= \pi (n_B - \overline{n}_B ) .
\end{eqnarray}
\end{mathletters}
Rewriting the Hamiltonian (\ref{Efetov_Larkin}) with $\theta_J$ and
its conjugate operator $n_J \equiv {1 \over \pi} \nabla \theta_B$,
we again obtain a Gaussian model,
\begin{eqnarray}
  \label{conjugate}
  {\cal H}_{\rm EL}
   &&= {\pi \over 2} \int dx
   \left\{
      v_J \bigl[ n_J (x) \bigr]^2
     + v_N
     \Bigl[
        { \nabla \theta_J (x) \over \pi }
     \Bigr] ^2
   \right\},
\end{eqnarray}
where $[n_J (x),\theta_J (x')]= i \delta (x-x')$.
This duality of the charge and current operators was also
emphasized in the Haldane's paper.\cite{haldane}  Since the
density-density correlation is written as an exponential of
$\theta_J$ in this representation, its calculation can be done
similarly as before,
\begin{eqnarray}
   N_B(r) &&\equiv \langle n_B (r) n_B (0) \rangle
   \approx {\overline{n}_B}^2 \cos (2k_F^{\rm SF} r )
   \langle e^{2i [ \theta_J (r) - \theta_J (0) ]} \rangle \nonumber \\
   &&\propto \cos (2k_F^{\rm SF} r ) \, r^{-K_{\rho}}.
  \label{density_corr}
\end{eqnarray}

Instead of calculating the sound velocity, the value of $K_\rho$ is
more accurately obtained through the Drude weight defined by
\begin{equation}
   D \equiv {L \over 2} \,
   {\partial^2 E_{\rm G.S.}(\phi )
    \over
    \partial \phi ^2 }
   \Bigg|_{\phi=0} \ ,
  \label{eq:Drude}
\end{equation}
where $E_{\rm G.S.}(\phi )$ is the total energy of the ground state
when the flux $\phi$ is penetrating.  Assuming the holes propagate
in pair, the relation between $D$ and $v_s$ for the ladder is
obtained and the result agrees with the one for the
chain\cite{schulz} aside from its numerical factor,
\begin{equation}
   D = {4 \over \pi} \, v_J = {2 \over \pi} \, K_\rho v_s \ .
  \label{eq:Drude_2}
\end{equation}
Combined this with Eq.\ (\ref{exponent}), we finally obtain the
formula of the correlation exponent in terms of the compressibility
and the Drude weight:
\begin{equation}
   K_\rho = \pi \, \rho \ \sqrt{\kappa D \over 2} .
  \label{eq:exponent_2}
\end{equation}
It is noted that this expression is identical to the one for the
chain in terms of $\kappa$ and $D$, while they differ by factor 2
in terms of $\kappa$ and $v_s$.

In this way, when the energy scale concerned is smaller than the
spin gap, we can predict long range asymptotic behavior of the
correlation functions based on the Efetov-Larkin effective model,
and calculate correlation exponents once the two parameters,
$v_s$ and $K_{\rm B}$ (or equivalently $\kappa$ and $D$),
are numerically determined.

Figure \ref{fig:K_rho} shows the correlation exponent
calculated in this way for the isotropic ladder ($J'/J=1$)
with $L=7$ and two holes as a function of $J$.  This corresponds
to the electron density, $\rho ={6 \over 7}=0.857$.  The necessary
quantities are carefully calculated by using the Lanczos
diagonalization.  The compressibility, $\kappa$, is determined
{}from the ground state energy of unfrustrated systems in the sense
explained before.  The Drude weight, $D$, is then calculated by
Eq.\ (\ref{eq:Drude}).  Combining these two, $K_\rho$ is obtained
via Eq.\ (\ref{eq:exponent_2}).

Near the phase separation boundary, $K_\rho$ grows rapidly.
This is owing to the divergence of the compressibility at this
boundary. In other words, the collective charge excitations become
softening and the superconducting fluctuations are enhanced
correspondingly.

Recently Hayward et al.\cite{hayward} directly calculated various
correlation functions for the $t$-$J$ ladders by using the
density-matrix renormalization-group method.  At $J/t = J'/t =1$ and
$\rho =0.8$, they found a power-law decay of the pairing
correlations with the exponent close to unity.  This exponent
corresponds to $K_\rho \sim 1$, which is larger than our estimate
at the same $J$ and $J'$, $K_\rho \approx 0.7$.\cite{tsunetsugu95}
The discrepancy may be owing to the larger electron density in our
calculation, $\rho = {6 \over 7}=0.857$, but it is not clear if
this suffices to account for the difference until new calculation
is carried out at the same density.  Hayward et al. also report
data on the density-density correlation function\cite{hayward} but
the exponent of the expected power-law decay of the oscillatory
term cannot be easily extracted.

\section{One-Particle Excitations}
\label{sec:chargexc}

Finally we discuss the one-particle Green's function where we can
see the quasiparticle excitations directly.  The electron and hole
parts of its spectral function are defined as
\begin{eqnarray}
  A_{\rm e,\sigma} ({\bf k},\omega)
  && \equiv \sum_n
    \Bigl| \langle n, \mbox{2$L$$-$1}
           | c_{{\bf k}\sigma}^\dagger
           |{\rm G.S.},\mbox{2$L$$-$2}
      \rangle
    \Bigr|^2
  \nonumber\\
  && \times
     \delta (
              \omega - E_n   (\mbox{2$L$$-$1})
                     + E_{\rm G.S.} (\mbox{2$L$$-$2}) + \mu
            ),
  \nonumber\\
  A_{\rm h,\sigma} ({\bf k},\omega)
  && \equiv \sum_n
    \Bigl| \langle n, \mbox{2$L$$-$3}
           | c_{{\bf k}\sigma}
           |{\rm G.S.},\mbox{2$L$$-$2}
      \rangle
    \Bigr|^2
  \nonumber\\
  && \times
     \delta (
              \omega + E_n   (\mbox{2$L$$-$3})
                     - E_{\rm G.S.} (\mbox{2$L$$-$2}) + \mu
            ),
\label{eq:spe}
\end{eqnarray}
where $|n , N \rangle$ is an eigenstate for $N$ electrons with the
energy $E_n (N)$ and $|{\rm G.S.},N\rangle$ denotes the ground state
with $N$ electrons.  Positive (negative) energies correspond to the
electron (hole) part.  The chemical potential is defined by
\begin{equation}
  \mu \equiv {\textstyle {1 \over 2}}
  \bigl[
     E_{\rm G.S.}(\mbox{2$L$$-$1})-E_{\rm G.S.}(\mbox{2$L$$-$3})
  \bigr].
\end{equation}

The results for $L=10$ and the isotropic couplings, $J'/J=1$, are
shown in Figs.~\ref{fig:GF} for $J/t$=0.3 and 0.5.  The wave
vectors along the ladder $k_x = {2 \pi \over L} n$ ($n$: integer)
are for PBC's and $k_x = {2 \pi \over L} (n + {1 \over 2})$ for
APBC's.  The ground state energy $E_{\rm GS}(2L-2)$ and the
chemical potential $\mu$ in Eq.\ (\ref{eq:spe}) are the average
over both boundary conditions.  The results for $L=10$ are very
similar to our $L=8$ results published before.\cite{tsunetsugu94}

The spectral function has large weights for the bonding ({\sl B})
($k_y =0$) and antibonding ({\sl A}) ($k_y = \pi$) orbitals only
near the Fermi energy $\omega = 0$, and they seem to constitute
quasiparticle bands.  Away from the Fermi energy, the individual
quasiparticle peaks are much less prominent and there is an
incoherent part with an energy of the order of $t$.

The quasiparticle part of the spectrum is consistent with the
mean-field theory based on the $d$-wave RVB state.\cite{sigrist}
The undoped ladder consists of local singlets on the rungs.
Such a singlet is the superposition of two electrons in the bonding
orbital and two electrons in the antibonding orbital,
$  b_{\uparrow}^{\dag} b_{\downarrow}^{\dag}
  -a_{\uparrow}^{\dag} a_{\downarrow}^{\dag} |0\rangle$.
Holes doped into the half-filled ladder will go predominantly into
the anti-bonding orbitals to gain a larger kinetic energy along the
rung direction.  The bonding band is occupied by more electrons,
while the antibonding band is occupied by less electrons.
The quasiparticle with energy closest to $\omega =0$ has a wave
vector nearest to the original Fermi wave number, $k_F$:
($k_x = {3\pi \over 5}$ for bonding and $k_x = {2\pi \over 5}$
for antibonding).  Because of the band splitting,
$k_F^{\rm B} > k_F^{\rm A}$, but the Luttinger sum rule is
satisfied, $k_F^{\rm B} + k_F^{\rm A} = (1-\delta) \pi $.
This means the Fermi volume is large, proportional to the electron
number rather than the hole number, and this is consistent with
photo-emission experiments on cuprate superconductors.\cite{dessau}
It is important to notice that the quasiparticle peaks near the
Fermi energy have their counterparts on the opposite side of the
Fermi energy.  An electronic quasiparticle peak at energy
$\omega >0$ has a shadow hole peak at energy around $-\omega <0$,
and {\it vice versa}.  These peaks indicate that the quasiparticle
excitations are those of the Bogoliubov quasiparticles as in BCS
theory, {\it i.e.}, mixture of an electron and a hole
($ \alpha_{\bf k}^\dagger = u_{\bf k} c_{{\bf k}
   \uparrow}^\dagger + v_{\bf k} c_{-{\bf k} \downarrow}$).
The weights in the electron and hole parts are proportional to
$|u_{\bf k}|^2$ and $|v_{\bf k}|^2$. They are hole-like around
$k_x =0$ and electron-like around $k_x =\pi$ for both the bonding
and antibonding bands.

There exists a finite energy gap in the quasiparticle spectra.
The electron and hole branches both come close to the Fermi energy
at $k_x \sim {\pi \over 2}$, but instead of passing through they
move away from it.  The energy gap for $J/t=0.3$ ($0.5$) is
$0.13t$ ($0.29t$) at ${\bf k}=({3\pi \over 5},0)$ and
$0.22t$ ($0.39t$) at ${\bf k}=({2\pi \over 5},\pi)$.
This corresponds to a quasiparticle gap
$2\Delta_{\rm QP} \simeq 0.13t \simeq {J \over 2}$ for
$J/t=0.3$ and $2\Delta_{\rm QP} \simeq 0.29t
 \simeq {3 J\over 5}$ for $J/t=0.5$.

It is interesting to note that the calculations of
$A({\bf k},\omega)$ in 2D clusters\cite{stephan,ohta} show similar
behavior for ${\bf k}$-points not along $(1,1)$ but no shadow peaks
for ${\bf k} \parallel (1,1)$, indicating $d_{x^2-y^2}$-pairing also.

Figure \ref{fig:DOS} shows the spectral function of the one-particle
Green's function of two holes in an $L=10$ ladder summed over all
wave vectors:
$A_{{\rm e},\sigma}(\omega)=
 \sum_{\bf k} A_{{\rm e},\sigma}({\bf k},\omega)$, and
$A_{{\rm h},\sigma}(\omega)=
 \sum_{\bf k} A_{{\rm h},\sigma}({\bf k},\omega)$.
This quantity is the local density of states to add and remove
electrons. In a strongly correlated system, the sum rules on the
weight (i.e., the integrated values) of $A_{\rm e}(\omega)$ and
$A_{\rm h}(\omega)$ are very different since the former is given by
the number of empty sites and the latter by the number of filled
sites (or equivalently the number of holes and electrons,
respectively).  In a Fermi liquid the values of $A_{\rm e}(\omega)$
and $A_{\rm h}(\omega)$ for small values of $|\omega -\mu |$ are
determined by quasiparticle weight at the Fermi energy and are
continuous.  It is interesting therefore to note that
Fig.~\ref{fig:DOS} shows approximately similar values for
$A_{\rm e}(\omega -\mu)$ and $A_{\rm h}(\mu-\omega)$
around $\omega \sim \mu$, but the sum rule on the total weight is
satisfied through the large weight in incoherent excitations in
$A_{\rm h}(\omega)$ at higher energies, $|\omega - \mu | > J.$
The strong correlation condition is reflected in the much smaller
total weight in $A_{\rm e} (\omega)$ which comes about through an
effective cut-off in energy on $A_{\rm e} (\omega)$.
In this respect the system in energy space is similar to a lightly
hole doped band insulator although as we discussed earlier the
location in ${\bf k}$-space of the coherent quasiparticle peaks
corresponds to a large Fermi surface to add and remove electrons.

The momentum distribution for electrons,
$n^{\rm e}_\sigma({\bf k}) \equiv
 \langle c_{{\bf k},\sigma}^{\dag}c_{{\bf k},\sigma} \rangle$,
and for holes,
$n^{\rm h}_\sigma({\bf k}) \equiv
 \langle c_{{\bf k},\sigma}c_{{\bf k},\sigma}^{\dag} \rangle$,
is shown in Fig.~\ref{fig:n_k}.  Note because of the strong
correlation condition, these do not add to one but instead their
sum is given by
$ n^{\rm e}_\sigma({\bf k}) +n^{\rm h}_\sigma({\bf k})
  = {1 \over 2}(1+\delta)$.
The strong correlation condition is also evident in the reduced
magnitude of the variation of $n^{\rm e}_\sigma({\bf k})$ as a
function of ${\bf k}$.  Nonetheless the presence of an apparent
``Fermi surface'' in the center of the Brillouin zone is clear,
consistent with the dispersion relations of the coherent
quasiparticles.  The difference between the bonding ($k_y=0$) and
antibonding ($k_y=\pi$) bands and the reduced occupation of the
antibonding band arise from the energy gain in placing the doped
holes preferentially in the antibonding band.

\section{Conclusions}

The results of our Lanczos diagonalizations confirm earlier studies
which concluded that lightly doped two-leg ladders belong to a
different universality class from single chains.  The latter are
Tomonaga-Luttinger liquids with gapless and separated spin and
charge excitations. The ladder in contrast has a finite gap in the
spin excitation spectrum and gapless excitations only in the charge
sector.  The low energy excitations evolve continuously from the
limit of strong inter-chain exchange coupling ($J' \gg J,t$) and
the simplicity of that limit allows a clear interpretation of
our results.

At large $J'$ the dispersion relation of a single doped hole
consists of two cosine bands corresponding to bonding and
antibonding states on a rung.  Lowering $J'$ to the isotropic limit
($J'/J=1$) and setting both $J,J'<t$ changes the dispersion
relation substantially.  The coherent parts of both bands are
centered at energies $\sim -1.5t$ but the width is $\sim J$ only.
The spin and charge components are still bound but more loosely and
the large magnetic polarizability of the spin background introduces
longer-range hoppings.  Remarkably the form of the bands resembles
the noninteracting band structure so that a photoemission
experiment which removes electrons would measure in effect a large
``Fermi surface'' with bonding and antibonding pieces. In this
regime the quasiparticle propagation is strongly influenced by
the coupling to magnetic excitations.

When two holes are added they bind together on a single rung at
large $J'$ and remain bound although the size of the bound hole
pair increases as $J'$ approaches $J \sim t/2$.  Moreover the
qualitative features of the density-density
${\cal N}({\bf q},\omega)$ and spin-spin structure factor
${\cal S}({\bf q},\omega)$, which are easy to understand at large
$J'$, remain similar as $J'$ approaches $J \sim t/2$.
${\cal N}({\bf q},\omega)$ near ${\bf q}=(0,0)$ is dominated by
the low energy mode associated with the motion of hole pairs.
At large $q_x$, ${\cal N}({\bf q},\omega)$ has a broad peak at high
energies ($\sim 4t$) similar to that found by Ohta, Eder and
Maekawa for 2D clusters.\cite{ohta}  They interpreted this as local
excitations of single holes in the magnetic cloud or spin
bag.\cite{eder}  The dynamical spin structure factor,
${\cal S}({\bf q},\omega)$, also resembles 2D clusters and not 1D
chains when we compare to the results of Tohyama, Horsch and
Maekawa.\cite{tohyama}  The major weight is at energies $\sim J$.
The spin gap evolves discontinuously upon doping through the new
quasiparticle excitations that can be made by breaking a hole pair
into two separate single holes.  However the major weight of the
spin excitations remains in the collective magnon mode whose
dispersion evolves continuously from the $\delta=0$ limit,
although it is influenced by the continuum of quasiparticle
excitations.

We have also investigated the one-particle spectral functions to
add and remove electrons from the two-hole ground state.
These show clearly the unusual nature of this ``Fermi liquid''.
When electrons are removed (or holes added), the spectral weight is
spread over a large energy region ($\sim 6t$), but the coherent
part is limited only to energies $\sim J$ below the Fermi energy
$\mu$.  The energy dispersion relations show a large apparent
Fermi surface for the coherent quasiparticles and which matches
onto a similar one for adding electrons at energies greater than
$\mu$.  These ${\bf k}$-space features resemble a metal with a
large Fermi surface.  The property that resembles a lightly hole
doped insulator is the energy dependence of the spectral weight to
add an electron.  This shows a low energy cut-off
($\sim \delta \cdot 6t$) similar to a lightly hole doped band
insulator.  The result is an intriguing duality between
metallic-like features in ${\bf k}$-space and lightly hole doped
insulating features in energy space.

The overall properties of the lightly doped ladder place it in the
Luther-Emery class rather than the Tomonaga-Luttinger class of 1D
systems. The low energy properties of Luther-Emery liquids are
described by interacting hard-core bosons as shown by Efetov and
Larkin.  In the present case the Efetov-Larkin bosons are bound
hole pairs.  Two features distinguish the $t$-$J$ ladder from the
usual Luther-Emery liquids arising from attractive interactions.
One is the $d$-wave character of the pairing and the second is the
presence of magnon excitations and limited quasiparticle
excitations.  Note the magnon excitations cannot be viewed as the
collective mode of quasiparticles since the latter vanish as
$\delta\rightarrow 0$.  The system is not a standard Fermi liquid,
but rather is a new and interesting mixture of a dilute attractive
Fermi gas in which the hole binding energy remains finite as
$\delta\rightarrow 0$, and a dense Fermi liquid with an apparent
large Fermi surface in $\bf k$-space.

Comparing the ladder with the results by Tohyama et al., we see that
the ladder is very different from the single chain but similar to 2D
clusters in many respects. Both in ladders and in 2D clusters
$d$-wave pairs are found down to small $J/t$. The dynamical charge
and spin structure factors look remarkably similar and the
single-particle spectral functions indicate the existence of
Bogoliubov quasiparticles with a finite superconducting gap.
Thus we are lead to the conjecture that the $t$-$J$ model on 2D
clusters is a doped RVB spin liquid showing $d$-wave pairing,
similar to the ladder.

\acknowledgements

We wish to thank M. Sigrist, F.C. Zhang, H. Monien, R. Noack,
D. Poilblanc, P. Prelovsek, D.J. Scalapino, S.R. White, and
D. W\"urtz for helpful discussions.
This work has been supported by the Swiss National Fund under grant
number NFP-304030-032833, by an internal grant of ETHZ and by the
Centro Svizzero di Calcolo Scientifico CSCS Manno.  The calculations
have been performed on the Cray Y-MP/464 of ETH Z\"urich and
on the NEC SX-3/24R of CSCS Manno.



\begin{table}
\caption[*]{Parameters for the fit of the lowest lying bands of the
one-hole spectra to a dispersion of the form of
Eq. (\protect{\ref{eq:fitspec}}).}
\begin{tabular}{cc|rrrr}
\phantom{A}$J/t$\phantom{A}&
\phantom{A}$k_y$\phantom{A}&
$\Delta E$\phantom{A}&
$\alpha_1$\phantom{A}&
$\alpha_2$\phantom{A}&
$\alpha_3$\phantom{AA}\\
\hline
 0.3 &  0  & $-$1.476 &    0.160 & 0.103 & $-$0.026\phantom{A}  \\
 0.3 &$\pi$& $-$1.417 & $-$0.192 & 0.134 &    0.025\phantom{A}  \\
\hline
 0.5 &  0  & $-$0.865 &    0.263 & 0.189 & $-$0.007\phantom{A}  \\
 0.5 &$\pi$& $-$0.790 & $-$0.311 & 0.225 & $-$0.011\phantom{A}
\end{tabular}
\label{tab:fitspec}
\end{table}

\begin{figure}
\caption{The $t$-$J$ ladder with two legs and $L$ rungs.
The couplings along the legs are $t$, $J$ and those along the rungs
$t'$, $J'$.}
\label{fig:geometry}
\end{figure}
\begin{figure}
\caption{Ground state energies for the $t$ ladder ($t$-$J$ ladder
with $J=J'=0$) with two holes. Results are shown for systems with
$L=2,3,\ldots,10$ rungs and periodic (PBC) as well as antiperiodic
(APBC) boundary conditions.  The ferromagnetic state always has the
lowest energy for $L\ge 4$ rungs.}
\label{fig:FM_engy}
\end{figure}
\begin{figure}
\caption{Real-space spin correlations for the singlet ground state of
the $t$ ladder with two holes.  $L=10$ and PBC's are used.  The two
ferromagnetic domains can clearly be seen.}
\label{fig:FM_corr}
\end{figure}
\begin{figure}
\caption{Graphical representation of the low-lying states of the
$t$-$J$ ladder in the strong coupling limit $J'\gg J,t$.
(a) The undoped case.  (b) One hole goes into either the bonding
orbital or the antibonding orbital on one rung.  (c) In the ground
state for two holes both holes are on the same rung.  (d) Scattering
states of two holes.  (e) At higher energies there is the triplet
excitation similar to the undoped ladder.}
\label{fig:strong_limit}
\end{figure}
\begin{figure}
\caption{Binding energy of two holes, spin gap and energy of the
triplet excitation away from the bound hole pair.  $J/t=0.3$ and
$0.3\le J'/t \le 3.0$.  The size of the ladder is $L=8$ rungs.}
\label{fig:gaps}
\end{figure}
\begin{figure}
\caption{Hole-hole correlation functions for the ground state of the
$t$-$J$ ladder with two holes.  $J/t=0.3$, and $J'/J$=1 and 10.
The size of the ladder is $L=10$ rungs and APBC's are used,
which have a lower ground state energy than PBC's.  The inset shows
the size of the bound hole pair, $\xi$, in the two-hole ground
state as a function of $J'$ for different ladder sizes.  }
\label{fig:hole_corr}
\end{figure}
\begin{figure}
\caption{The line of phase separation in the $t$-$J$ ladder
determined from the coupling at which the compressibility diverges.}
\label{fig:phasesep}
\end{figure}
\begin{figure}
\caption{The nine different states for a single rung.}
\label{fig:nine_basis}
\end{figure}
\begin{figure}
\caption{Energy spectra for the $t$-$J$ ladder doped with one hole.
The case of large $J'$ ($J'/J=10$): (a) $J/t=0.3$ and (b) $J/t=0.5$
with $L=8$.  The isotropic case ($J'/J=1$): (c) $J/t=0.3$ and
(d) $J/t=0.5$ with $L=10$.  The results for $k_x={n\over L}\pi$
with even $n$ are for PBC's and with odd $n$ for APBC's. }
\label{fig:disp_1h}
\end{figure}
\begin{figure}
\caption{Fit of the lowest lying bands of the one-hole spectra to
the form of Eq. (\protect{\ref{eq:fitspec}}). (a) $J/t=J'/t=0.3$ and
(b) $J/t=J'/t=0.5$. The size is $L=10$.}
\label{fig:disp_fit}
\end{figure}
\begin{figure}
\caption{The hole-spin correlations for the ground state of the
ladder doped with one hole at $J'/J=1$and $10$. The ladder has
$L=10$ rungs and $J/t=0.3$. The ground state has $S^z =1/2$.}
\label{fig:hole-spin}
\end{figure}
\begin{figure}
\caption{Energy spectra for the $t$-$J$ ladder doped with two holes
for a large $J'$ ($J'/J=10$): $J/t=0.3$ with (a) PBC's and
(b) APBC's, and $J/t=0.5$ with (c) PBC's and (d) APBC's.  The size
of the ladder is $L=8$ rungs.  The states are classified according
to total spin $S$ and parity.}
\label{fig:disp_2h_strong}
\end{figure}
\begin{figure}
\caption{Energy spectra for the $t$-$J$ ladder doped with two holes
at the isotropic point ($J'/J=1$): $J/t=0.3$ with (a) PBC's and
(b) APBC's, and $J/t=0.5$ with (c) PBC's and (d) APBC's.  The size
of the ladder is $L=10$ rungs.  The lines are only guides for the
eye and do not necessarily connect related states.}
\label{fig:disp_2h_iso}
\end{figure}
\begin{figure}
\caption{Hole-hole [$(a)$ and $(c)$] and spin-hole [$(b)$ and $(d)$]
correlation functions for the two triplet excitations.  $L=8$.
Dashed lines are for the lowest triplet state and dashed-dotted
lines for the lowest state with non-vanishing weight in ${\cal S}
({\bf q}=(\pi,\pi),\omega)$.}
\label{fig:corrs}
\end{figure}
\begin{figure}
\caption{Dynamical spin structure factor,
${\cal S}({\bf q},\omega)$, for the undoped ladder with $L=10$
rungs.  Note that the scale is different for $q_y=\pi$.}
\label{fig:spin_0h}
\end{figure}
\begin{figure}
\caption{Dynamical spin structure factor,
${\cal S}({\bf q},\omega)$, for the $L=10$ ladder with two holes.
$J/t=J'/t=0.3$ (upper panels) and $0.5$ (lower panels).
PBC's are used.  Note that the scale is different for around
${\bf q}=(\pi,\pi)$.}
\label{fig:spin_2h}
\end{figure}
\begin{figure}
\caption{Dynamical charge structure factor,
${\cal N}({\bf q},\omega)$, for the $L=10$ ladder with two holes.
$J/t=J'/t=0.3$ (upper panels) and $0.5$ (lower panels).
PBC's are used.}
\label{fig:charge_2h}
\end{figure}
\begin{figure}
\caption{Superconducting order parameter, $\chi_{\bf k}$ for
$N_{\rm h}=2$ calculated on the $L=8$ ladder with PBC's for several
values of $J/t = J'/t$.}
\label{fig:pair_k}
\end{figure}
\begin{figure}
\caption{The order parameter for the optimal pair
Eq.\ (\protect{\ref{eq:optimal}}) as a function of $J/t$ on an $L=8$
ladder with both PBC and APBC. Shown are results for 2 and 4 holes.}
\label{fig:OP}
\end{figure}
\begin{figure}
\caption{Correlation exponent, $K_\rho$, for the $L=7$ ladder with
two holes.}
\label{fig:K_rho}
\end{figure}
\begin{figure}
\caption{
Spectral function of the one-particle Green's function,
$A({\bf k},\omega)$, for the $L=10$ ladder with two holes.
(a) $J/t = J'/t = 0.3$ and (b) $J/t = J'/t =0.5$.
The width of each line represents the strength of the excitation.
For $\omega>0$ we show the spectral function for {\it adding} one
electron $A_{{\rm e},\sigma}({\bf k},\omega)$, and for $\omega<0$
the spectral function for {\it removing} an electron
$A_{{\rm h},\sigma}({\bf k},\omega)$.}
\label{fig:GF}
\end{figure}
\begin{figure}
\caption{Spectral function of the one-particle Green's function for
two holes summed over all wave vectors; $A_{{\rm h},\sigma}(\omega)$
=$\sum_{\bf k}A_{h,\sigma}({\bf k},\omega)$ and
$A_{{\rm e},\sigma}(\omega)$=
$\sum_{\bf k}A_{{\rm e},\sigma}({\bf k},\omega)$.  $L=10$ and
$J/t = J'/t = 0.3$.  The oscillations at large $|\omega|$ are caused
by nonconvergent Lanczos iterations at these energies.}
\label{fig:DOS}
\end{figure}
\begin{figure}
\caption{The momentum distribution function for the ground state for
$\delta=0.1$ and $L=10$.  $J/t = J'/t = 0.3$.  The electron part,
$n^{\rm e}_\sigma({\bf k})$=
$\langle c_{{\bf k},\sigma}^{\dag}c_{{\bf k},\sigma} \rangle$,
and the hole part,
$n^{\rm h}_\sigma({\bf k})$=
$\langle c_{{\bf k},\sigma}c_{{\bf k},\sigma}^{\dag} \rangle$.
The momenta, $k_x= 2\pi n /L$, with integer $n$'s are for PBC's,
while those with half-integer $n$'s are for APBC's.}
\label{fig:n_k}
\end{figure}


\begin{references}

\bibitem[*]{byline} Current address: Institute for Solid State Physics,
University of Tokyo, Roppongi 7-22-1, Tokyo 106, Japan

\bibitem{dagotto}        
E. Dagotto, J. Riera, and D. J. Scalapino,
Phys. Rev. B {\bf 45}, 5744 (1992).

\bibitem{noack}          
R.M. Noack, S.R. White and D.J.  Scalapino,
Phys. Rev. Lett. {\bf 73}, 882 (1994).

\bibitem{tsunetsugu94}   
H. Tsunetsugu, M. Troyer and T.M. Rice,
Phys. Rev. B {\bf 49}, 16078 (1994).

\bibitem{rice}           
T. M. Rice, S. Gopalan, and M. Sigrist,
Europhys. Lett. {\bf 23}, 445 (1993).

\bibitem{sigrist}        
M. Sigrist, T. M. Rice, and F. C. Zhang,
Phys. Rev. B {\bf 49}, 12058 (1994).

\bibitem{khveshchenko}   
D. V. Khveshchenko and T. M. Rice,
Phys. Rev. B {\bf 50}, 252 (1994);
D. V. Khveshchenko,
{\it ibid.} {\bf 50}, 380 (1994);
N. Nagaosa and M. Oshikawa, preprint cond-mat/9412003;
H. J. Schulz, preprint cond-mat/9412098.

\bibitem{tsunetsugu95}   
H. Tsunetsugu, M. Troyer and T.M. Rice,
Phys. Rev.  B {\bf 51}, 16456 (1995).

\bibitem{hirsch88}       
R. Hirsch, Diplomarbeit Universit\"at K\"oln, 1988.

\bibitem{strong}         
S. P. Strong and A. J. Millis,
Phys. Rev. Lett. {\bf 69}, 2419 (1992).

\bibitem{gopalan}        
S. Gopalan, T.M. Rice and M. Sigrist,
Phys. Rev. B {\bf 49}, 8901 (1994).

\bibitem{barnes93}       
T. Barnes, E. Dagotto, J. Riera and E.S. Swanson,
Phys. Rev.  B {\bf 47}, 3196 (1993).

\bibitem{barnes94}       
T. Barnes and J. Riera,
Phys. Rev. B {\bf 50}, 6817 (1994).

\bibitem{white}          
S.R. White and R.M. Noack and D.J. Scalapino,
Phys. Rev. Lett. {\bf 73}, 886 (1994).

\bibitem{troyer}         
M. Troyer, H. Tsunetsugu and D. W\"urtz,
Phys. Rev. B, {\bf 50}, 13515 (1994).

\bibitem{takano}         
M. Takano et al.,
JJAP Series {\bf 7}, 3 (1992);
M. Azuma, Z. Hiroi, M. Takano, K. Ishida, Y. Kitaoka,
Phys. Rev. Lett. {\bf 73}, 3463 (1994).

\bibitem{johnston}       
D. C. Johnston et al.,
Phys. Rev. B {\bf 35}, 219 (1987).

\bibitem{ogata_A}        
M. Ogata, M.U. Luchini and T.M. Rice,
Phys. Rev. B {\bf 44}, 12083 (1991).

\bibitem{imada}          
M. Imada,
Phys. Rev. B {\bf 48}, 550 (1993), and references therein.

\bibitem{reigrotzki}     
M. Reigrotzki, H. Tsunetsugu and T.M. Rice,
J. Phys: Cond.  Matt. {\bf 6}, 9235 (1994);
In Ref.~\onlinecite{barnes93} a term
was omitted in the strong coupling expansion.

\bibitem{nagaoka}        
Y. Nagaoka, Phys. Rev. {\bf 147}, 392 (1966);
D.J. Thouless, Proc. Phys. Soc. {\bf 86}, 893 (1965).

\bibitem{doucot}         
B. Doucot and X.G. Wen,
Phys. Rev. B {\bf 40}, 2719 (1989).

\bibitem{linden}         
W. van der Linden and D.M. Edwards,
J. Phys. Condens.  Matter {\bf 3}, 4917 (1991);
{\it ibid.} {\bf 3}, 7229 (1991);
A.J. Basile and V. Elser,
Phys. Rev. B {\bf 41}, 4842 (1990);
B.S. Shastry, H.R. Krishnamurthy, and P.W. Anderson,
Phys. Rev. B {\bf 41}, 2375 (1990).

\bibitem{putikka}        
W.O. Putikka, M.U. Luchini and T.M. Rice,
Phys. Rev. Lett.  {\bf 68}, 538 (1992);
W.O. Putikka, M.U. Luchini and M. Ogata,
{\it ibid.} {\bf 69}, 2288 (1992).

\bibitem{hirsch_phd}     
R. Hirsch, PhD thesis, Universit\"at K\"oln;
E.~M\"uller-Hartmann, private communications.

\bibitem{luther_A}       
A. Luther and V.J. Emery,
Phys. Rev. Lett. {\bf 33}, 589 (1974);
V.J. Emery,
in {\it Highly Conducting One-Dimensional Solids},
edited by J.T. Devreese et al. (Plenum, New York, 1979).

\bibitem{haldane}        
F.D.M. Haldane,
Phys. Rev. Lett. {\bf 45}, 1358 (1980);
J. Phys. C {\bf 14}, 2585 (1981).

\bibitem{ogata_B}        
M. Ogata, M.U. Luchini, S. Sorella and F.F. Assaad,
Phys. Rev. Lett. {\bf 66}, 2388 (1991).

\bibitem{K_takano}       
Similar excitations are also found for a $t$-$J$ chain
with frustrated couplings:
K. Takano and K. Sano,
Phys. Rev. B {\bf 48}, 9831 (1993);
I. Bose and S. Gayen,
Phys. Rev. B {\bf 48}, 10653 (1993).

\bibitem{heine}          
V. Heine in {\it Solid State Physics}, Vol. 35,
edited by H. Ehrenreich, F. Seitz and D. Turnbull
(Academic Press, New York, 1980)

\bibitem{eder}           
R. Eder, Y. Ohta and S. Maekawa,
Phys. Rev. Lett. {\bf 74}, 5124 (1995).

\bibitem{tohyama}        
T. Tohyama, P. Horsch and S. Maekawa,
Phys. Rev. Lett. {\bf 74}, 980 (1995).

\bibitem{efetov}         
K.B. Efetov and A.I. Larkin,
Zh. Eksp. Teor. Fiz. {\bf 69}, 764 (1975)
[Sov. Phys. JETP {\bf 42}, 390 (1976)].

\bibitem{mattis}         
D. C. Mattis and E. H. Lieb,
J. Math. Phys. {\bf 6}, 304 (1965).

\bibitem{luther_B}       
A. Luther and I. Peschel,
Phys. Rev. B {\bf 9}, 2911 (1974);
ibid. B {\bf 12}, 3908 (1975)

\bibitem{schulz}         
H. J. Schulz, Phys. Rev. Lett. {\bf 64}, 2831 (1990);
Int. J. Mod. Phys. B {\bf 5}, 57 (1991);
B. S. Shastry and B. Sutherland,
Phys. Rev. Lett. {\bf 65}, 243 (1990).

\bibitem{hayward}        
C. A. Hayward et al.,
{\it Evidence for a superfluid density in $t$-$J$ ladders},
preprint cond-mat/9504018.

\bibitem{dessau}         
D. S. Dessau et al.,
Phys. Rev. Lett. {\bf 71}, 2781 (1993).

\bibitem{stephan}        
W. Stephan and P. Horsch,
Phys. Rev. Lett. {\bf 66}, 2258 (1991).

\bibitem{ohta}           
Y. Ohta, T. Shimozato, R. Eder and S. Maekawa,
Phys. Rev. Lett. {\bf 73}, 324 (1994).

\end{references}
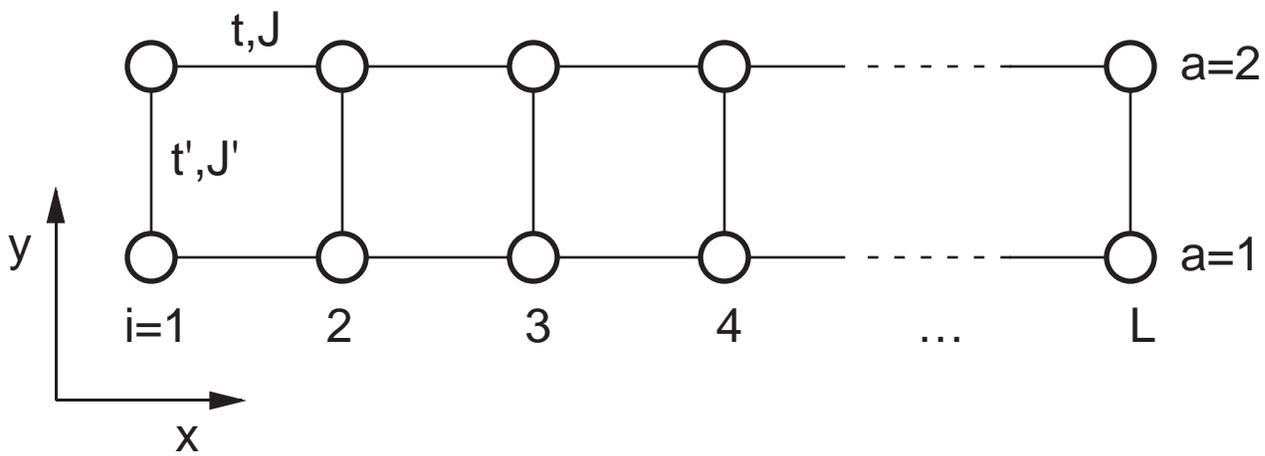
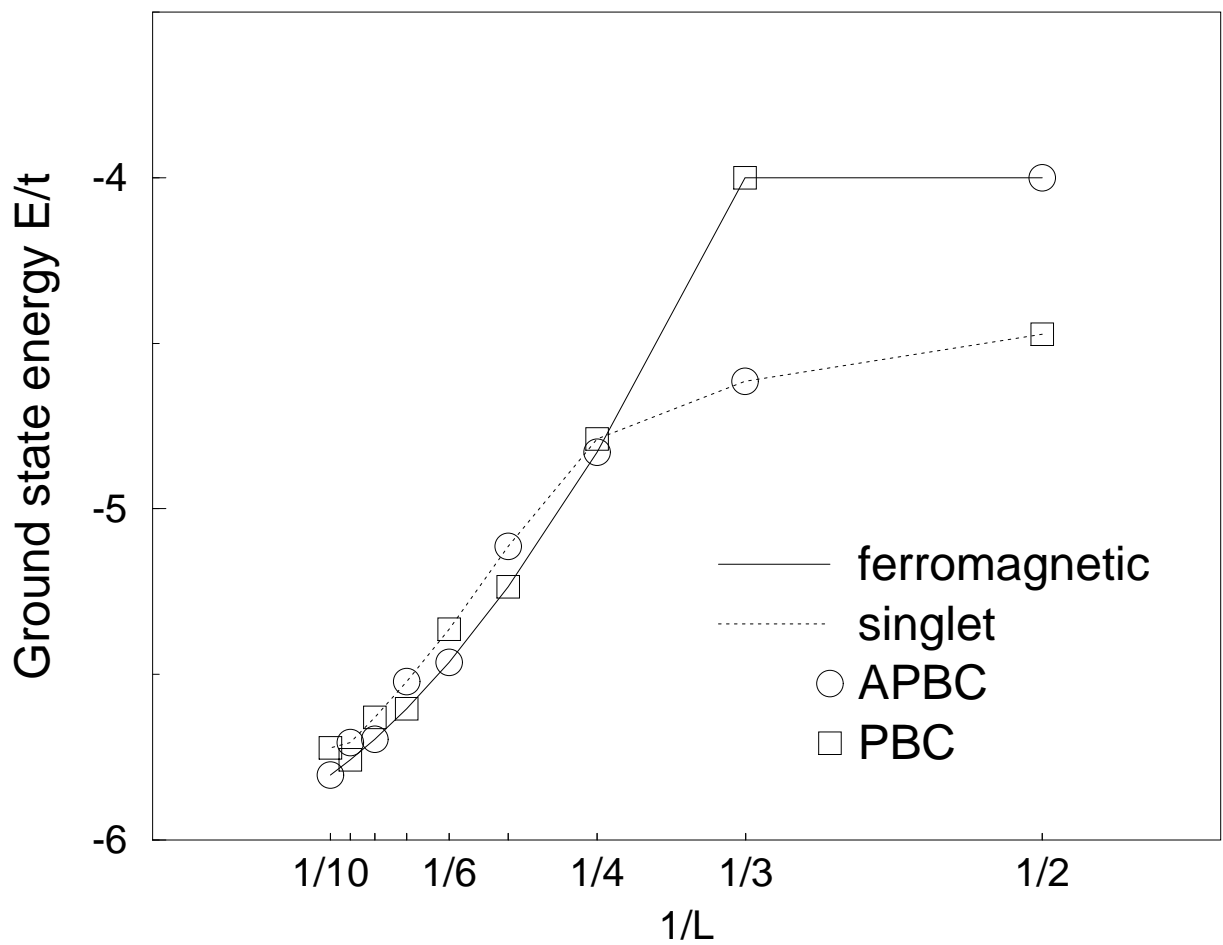
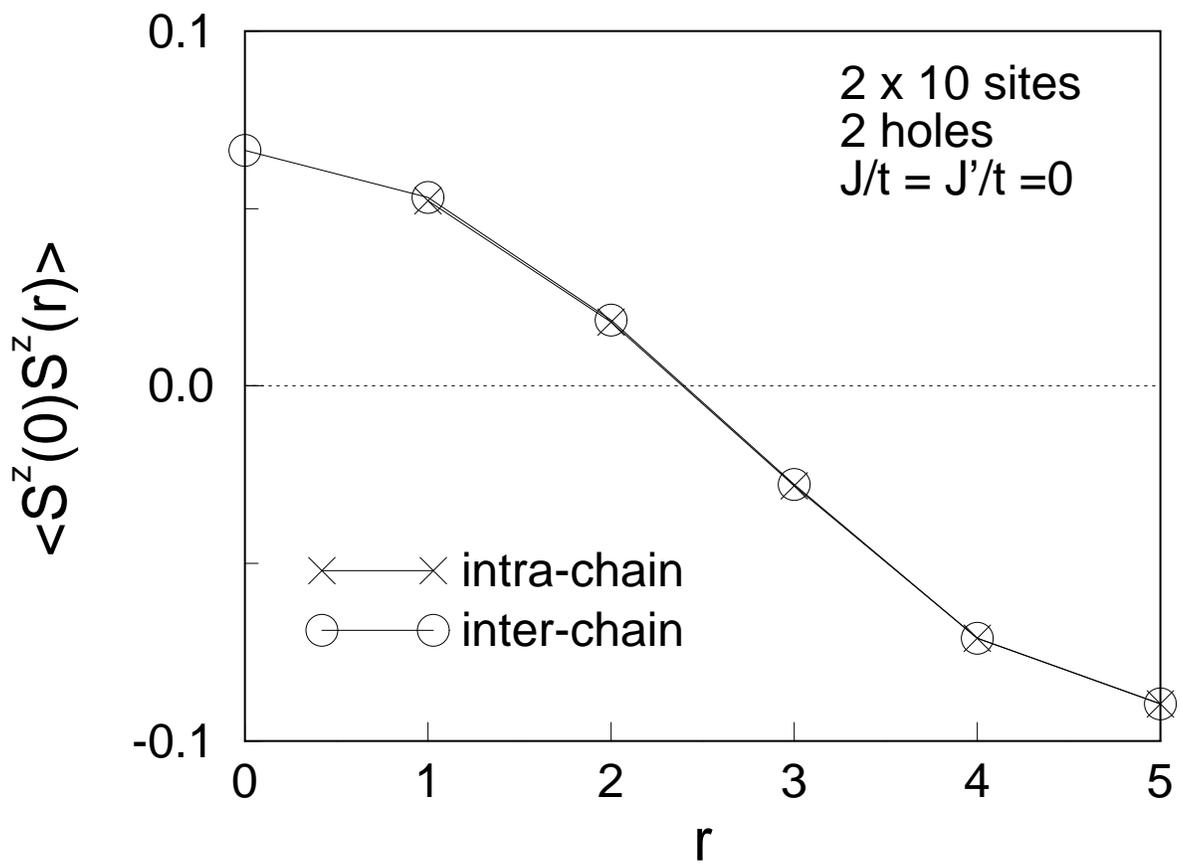
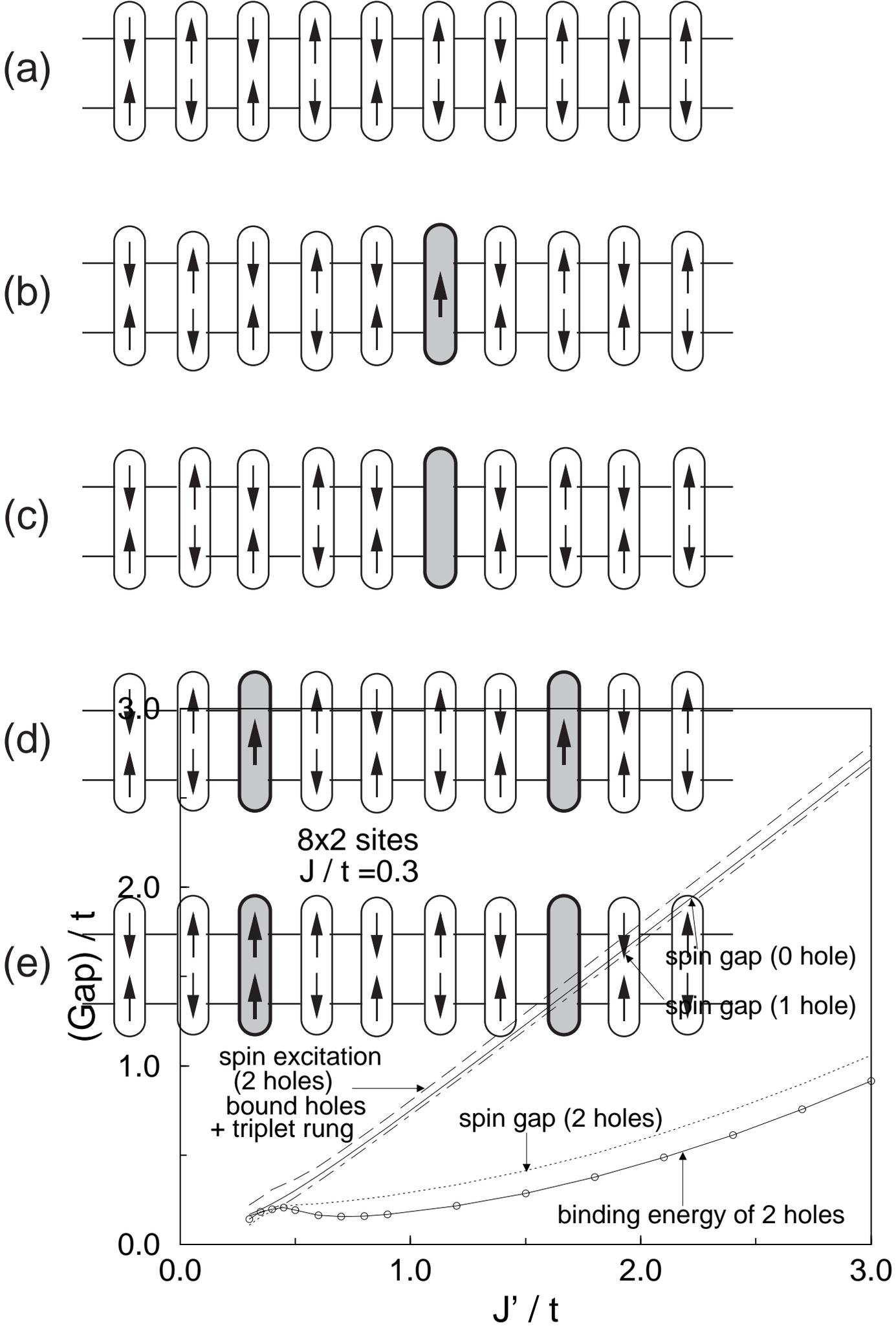
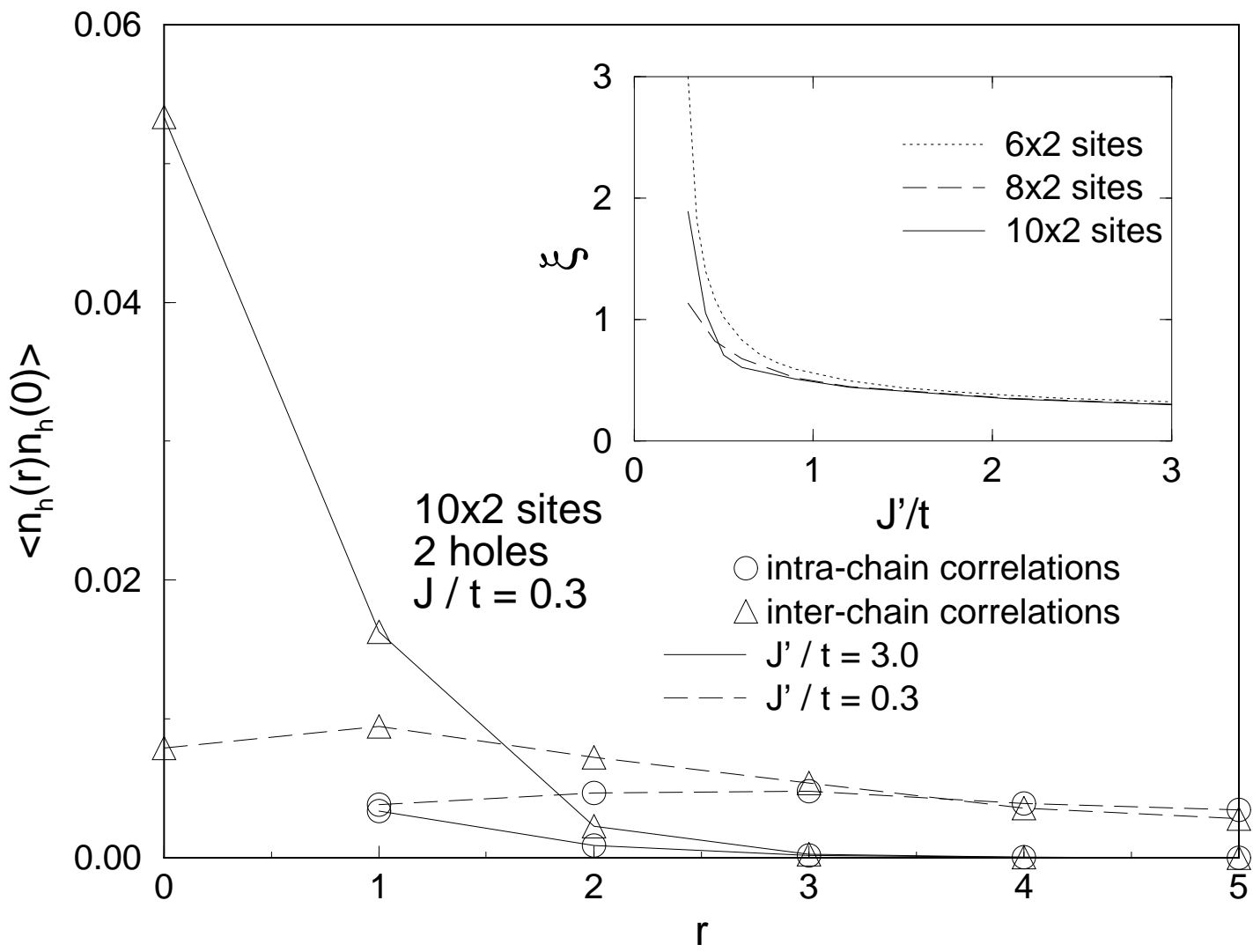
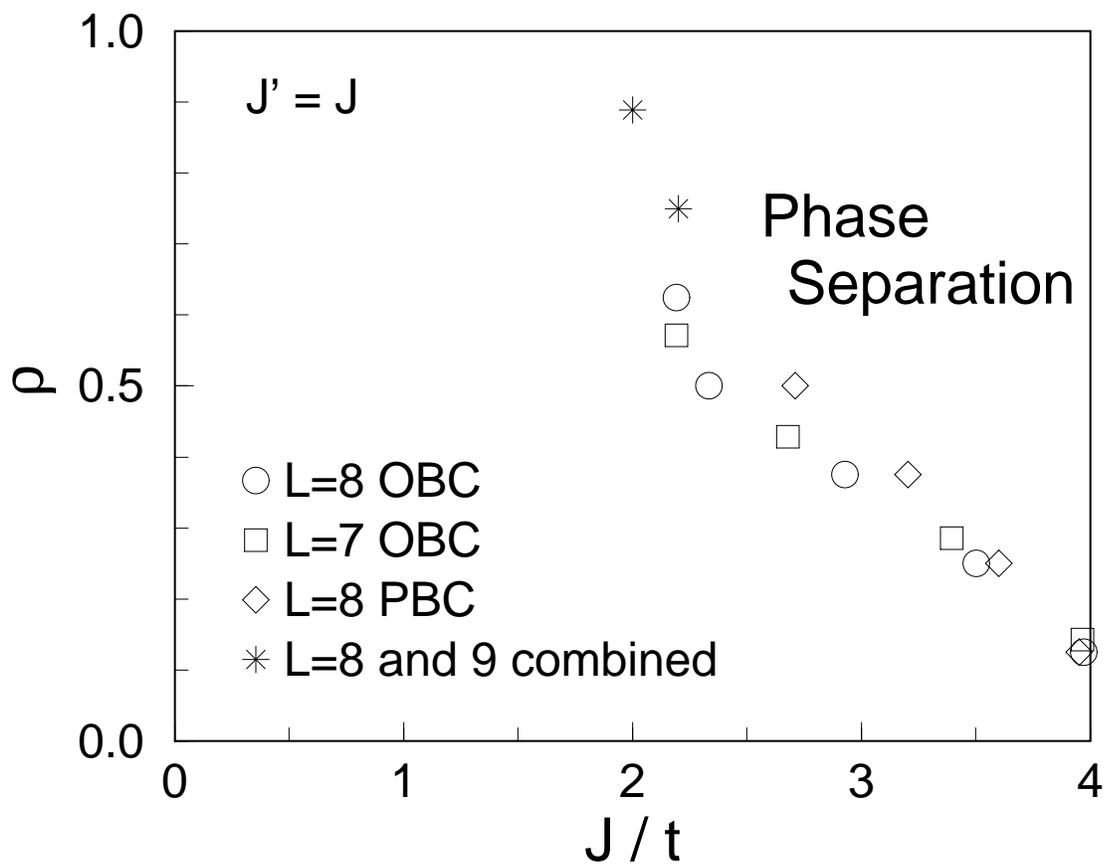
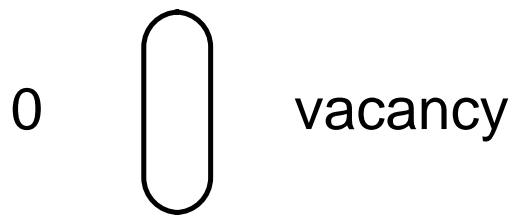
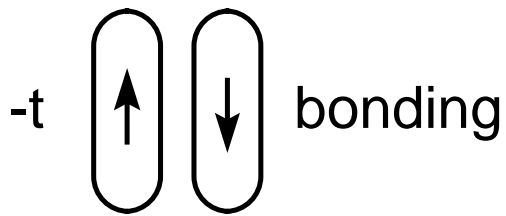
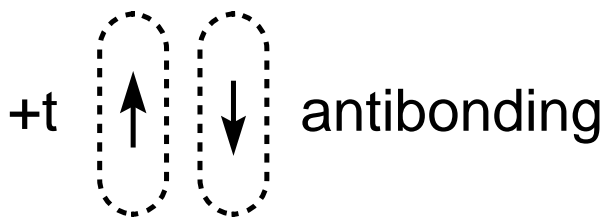
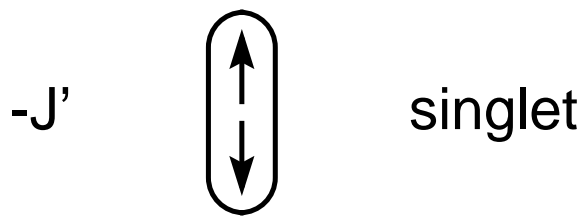
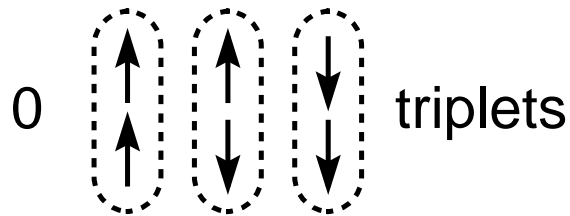
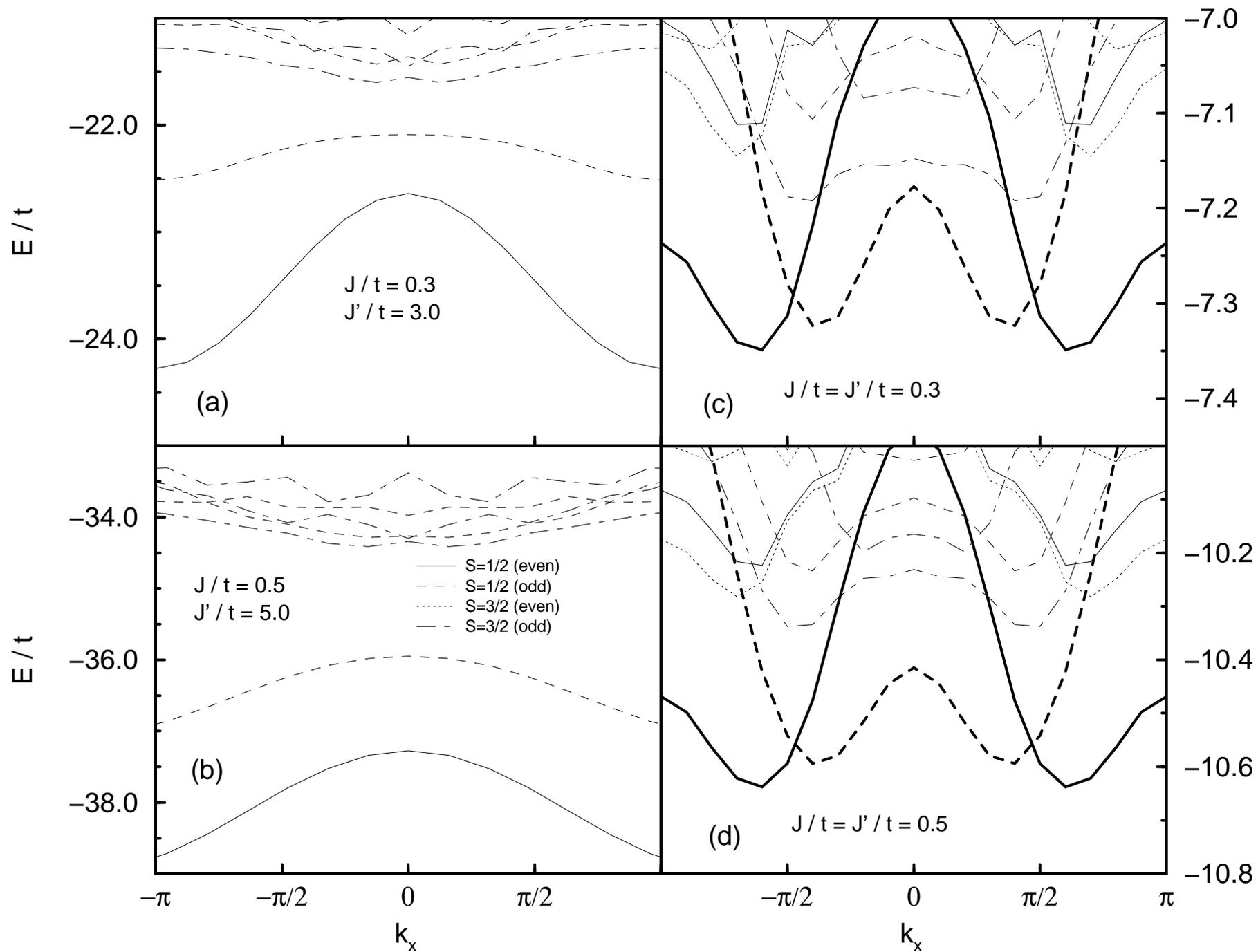
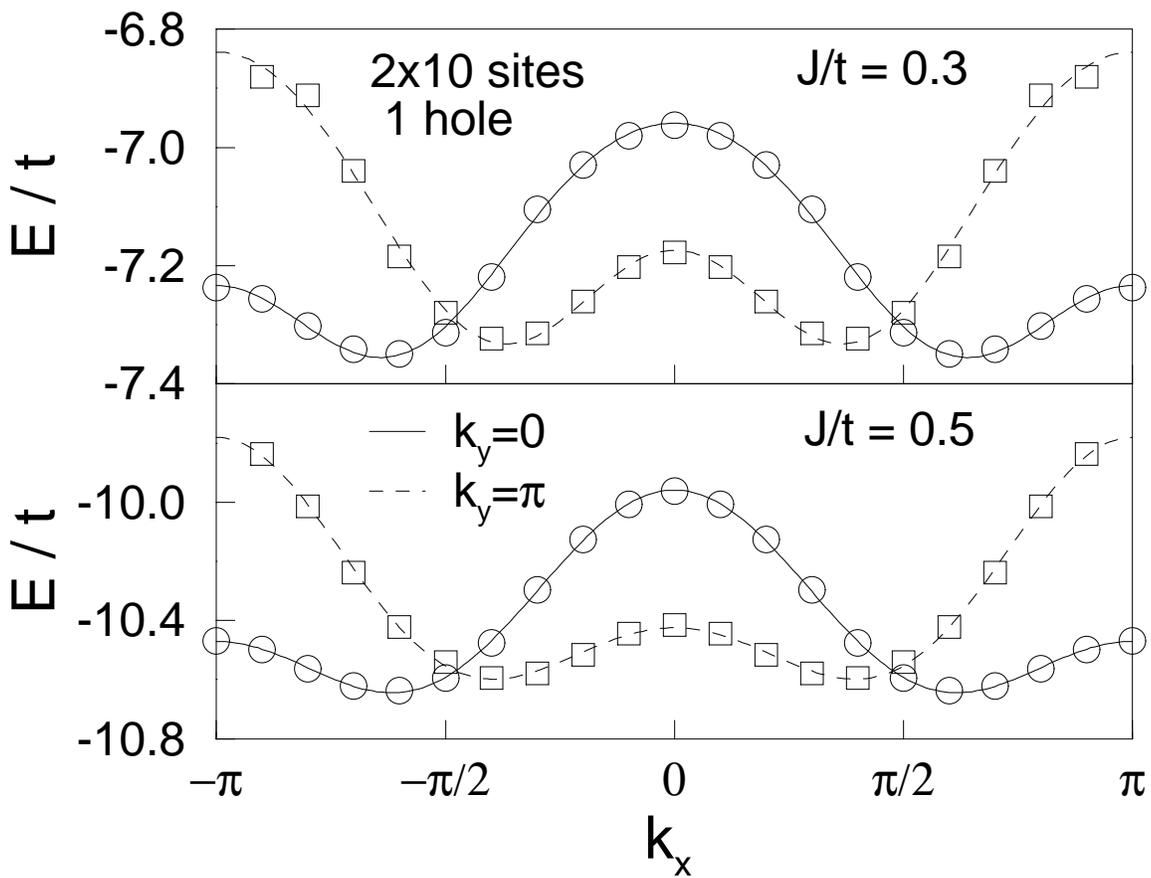
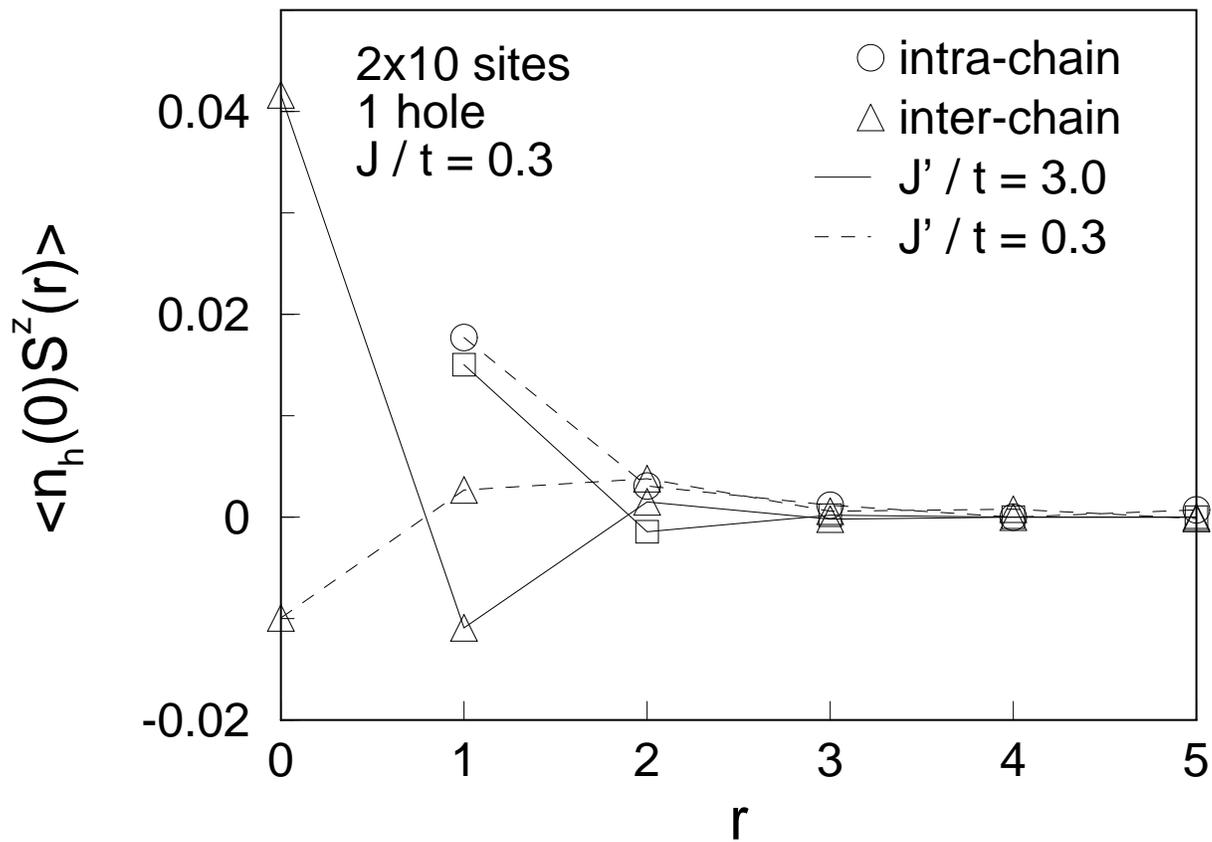
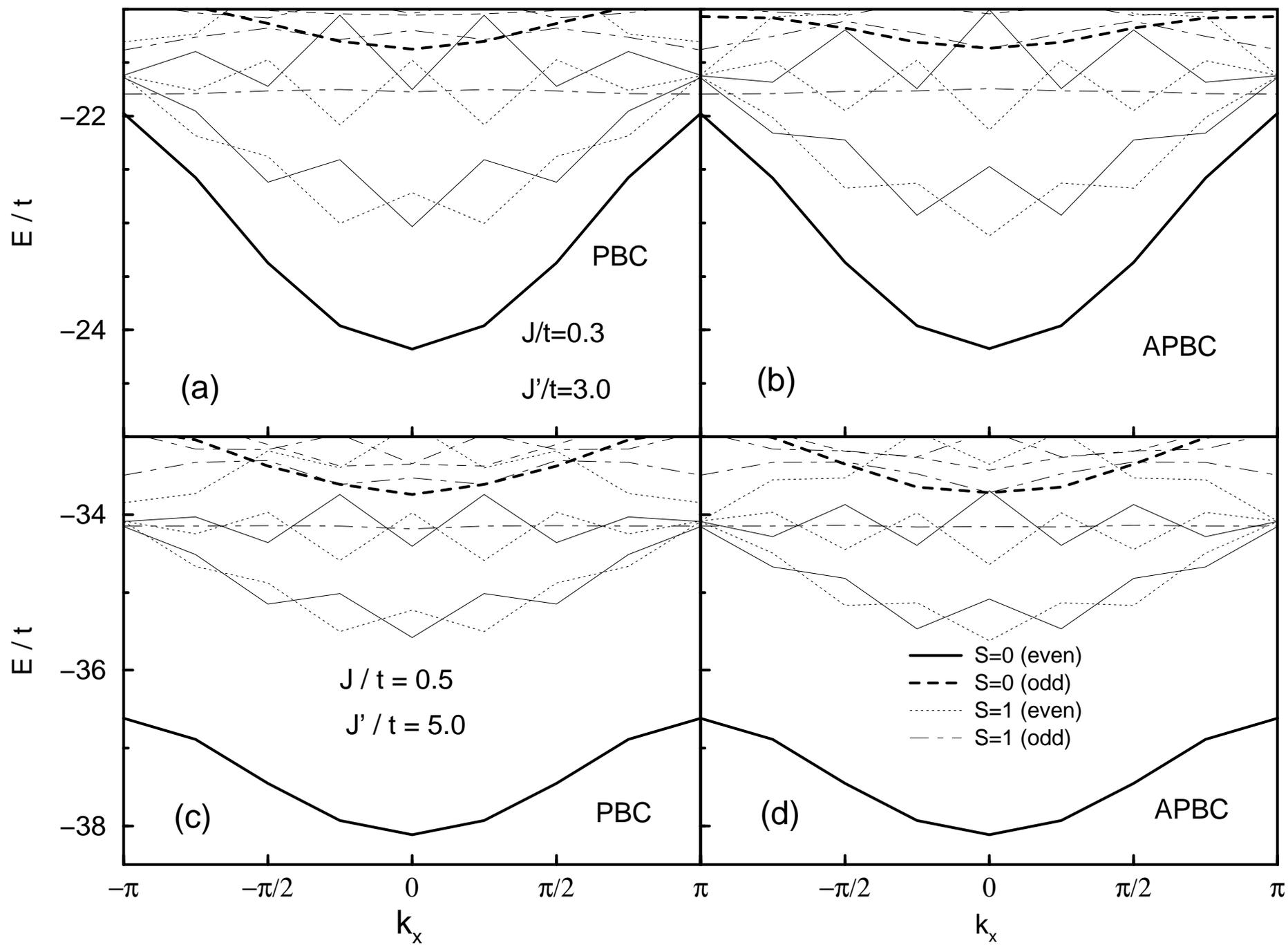
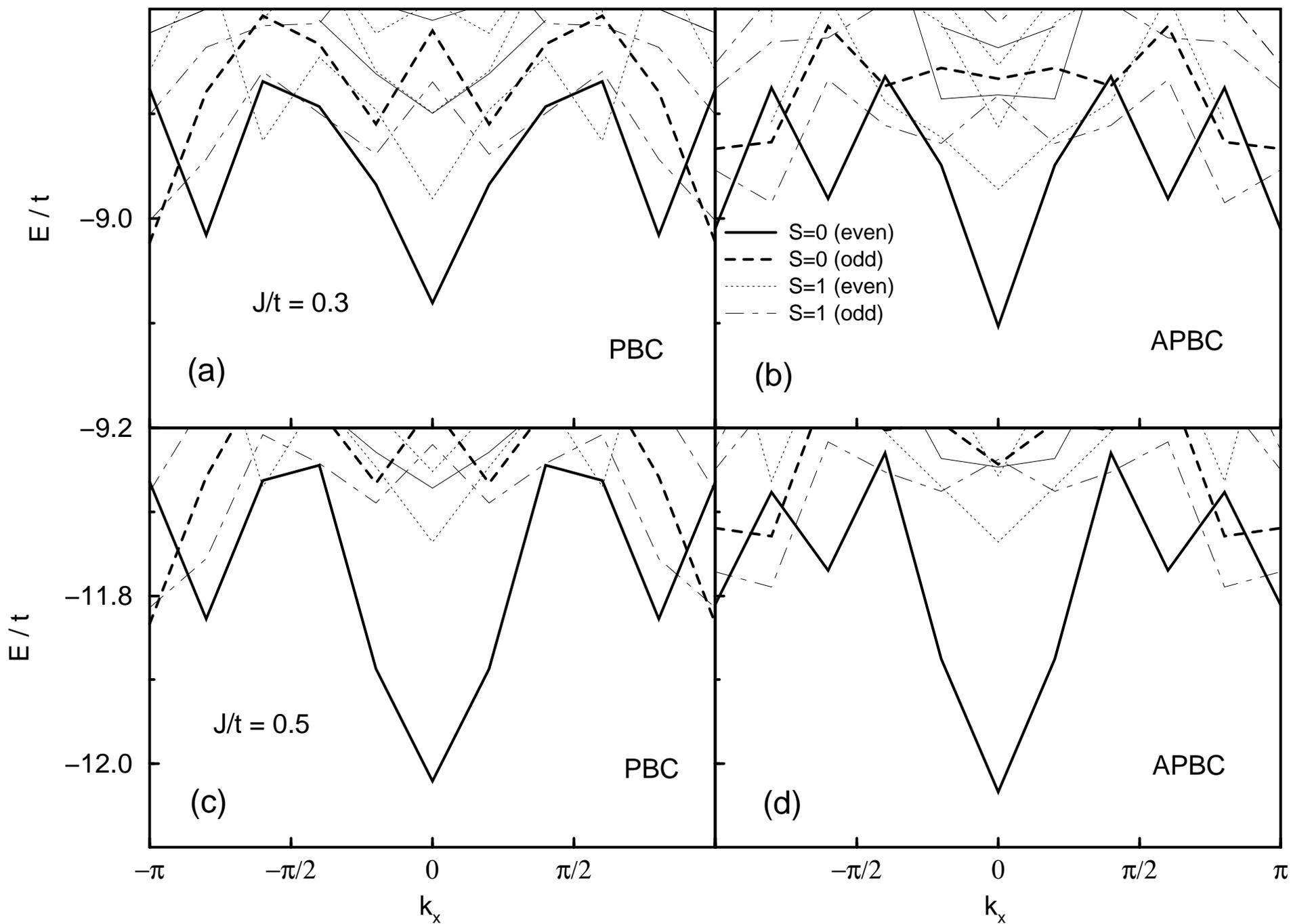
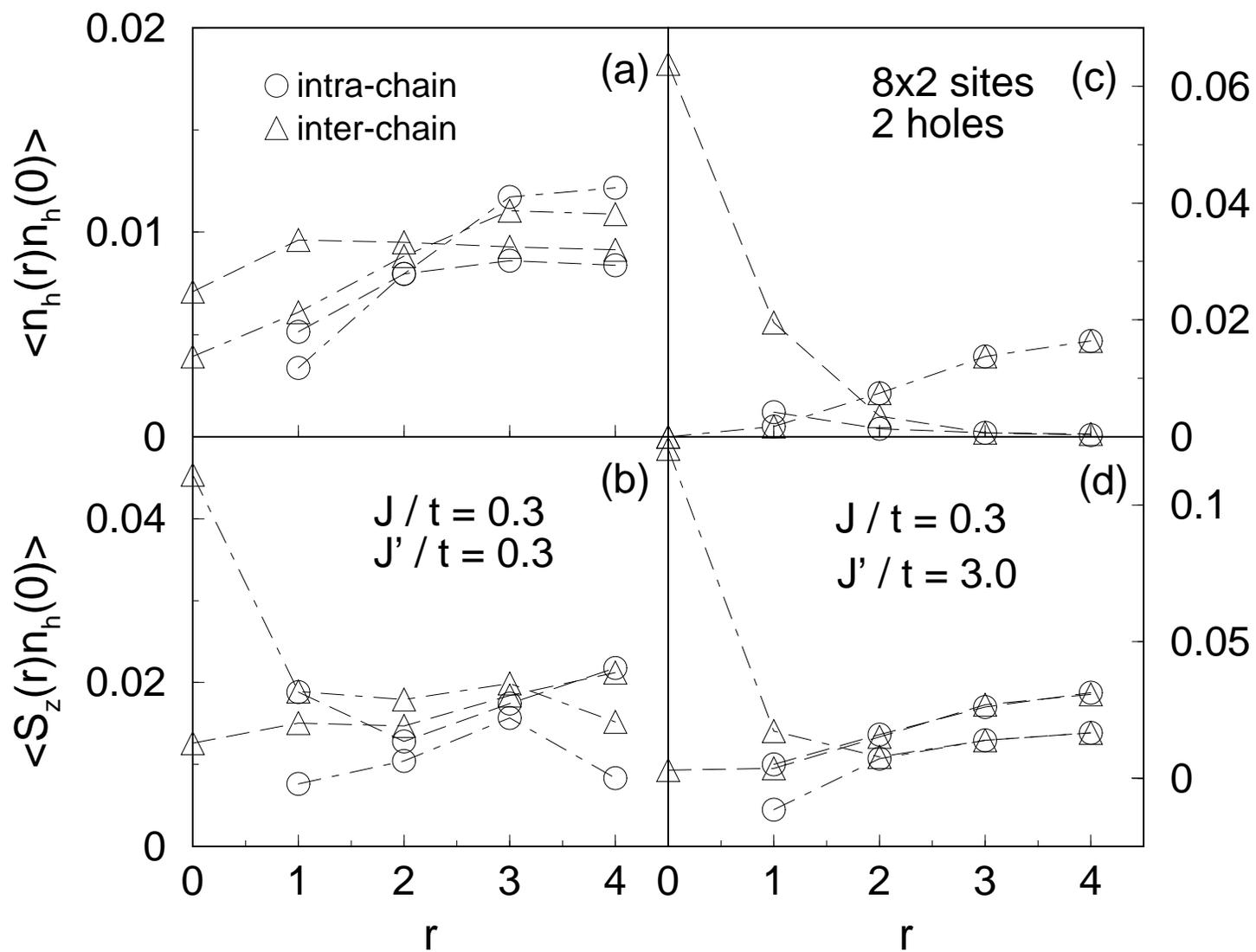
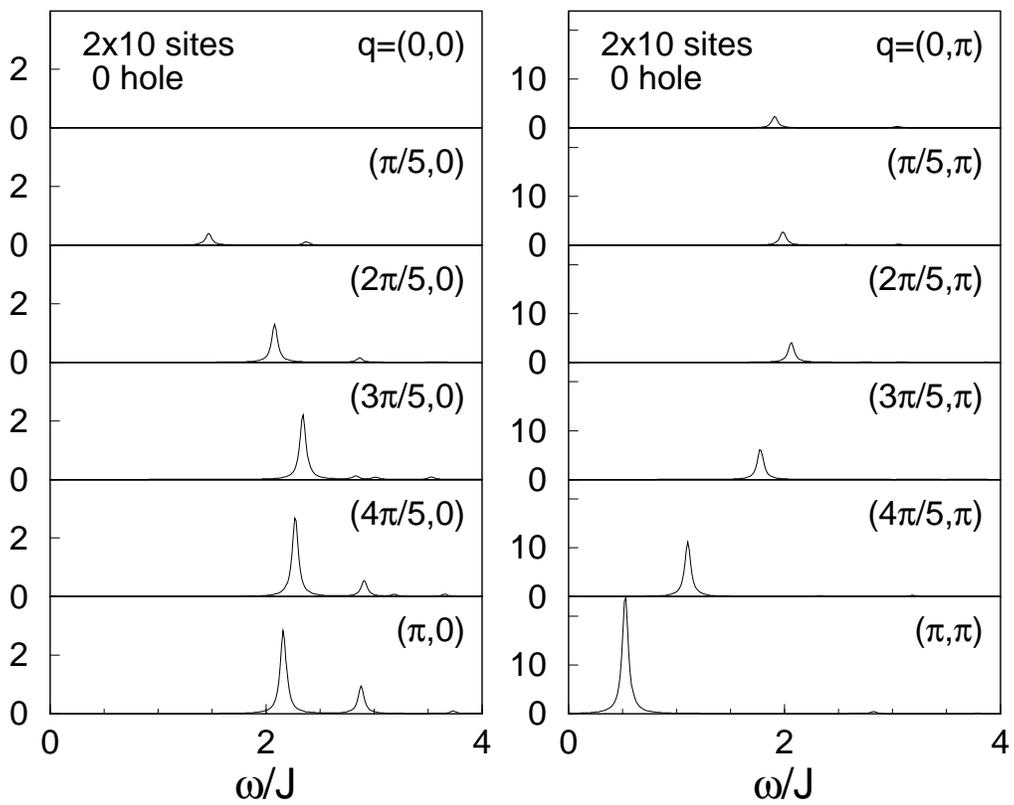
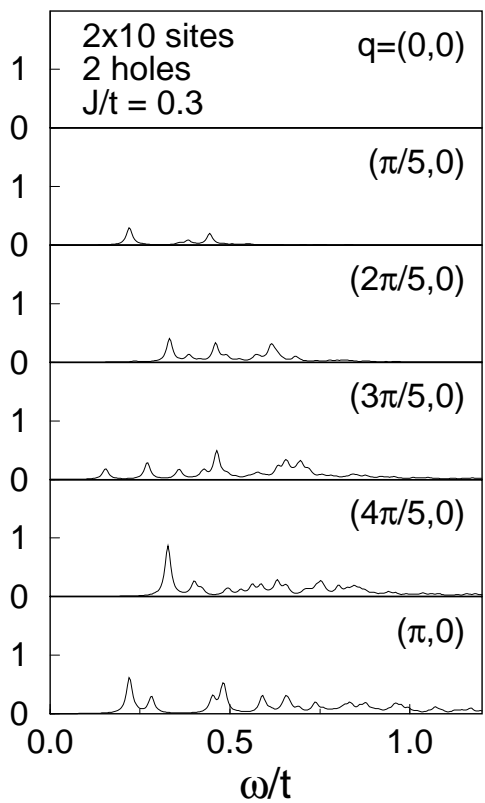
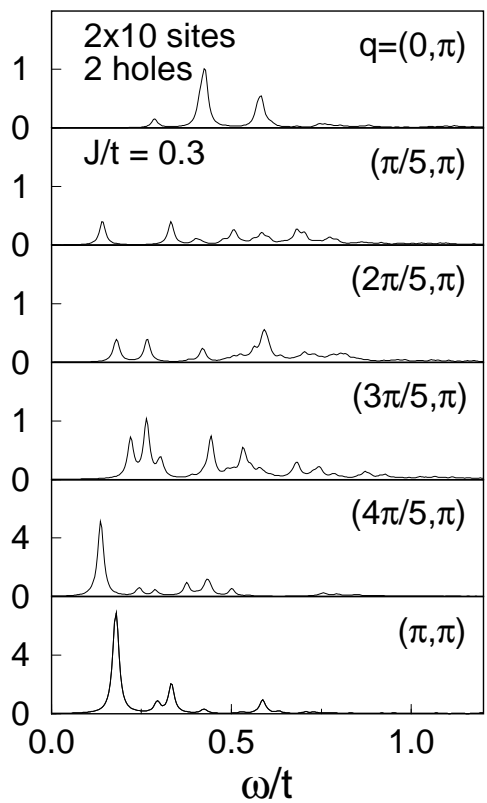
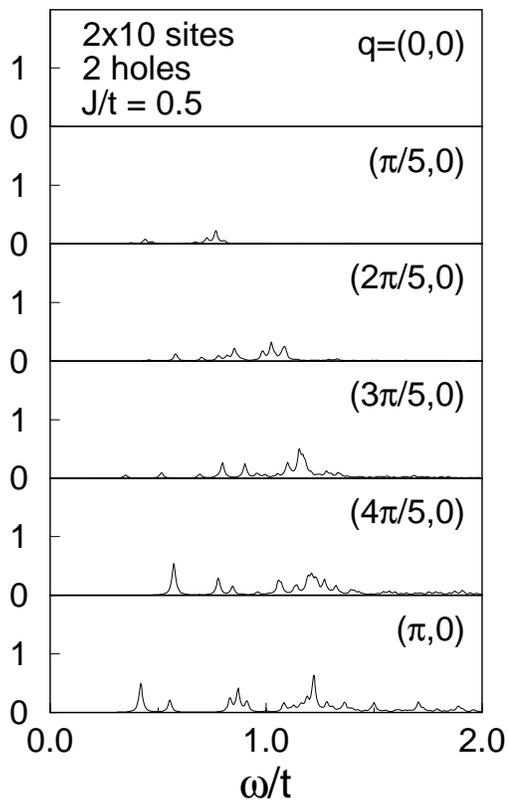
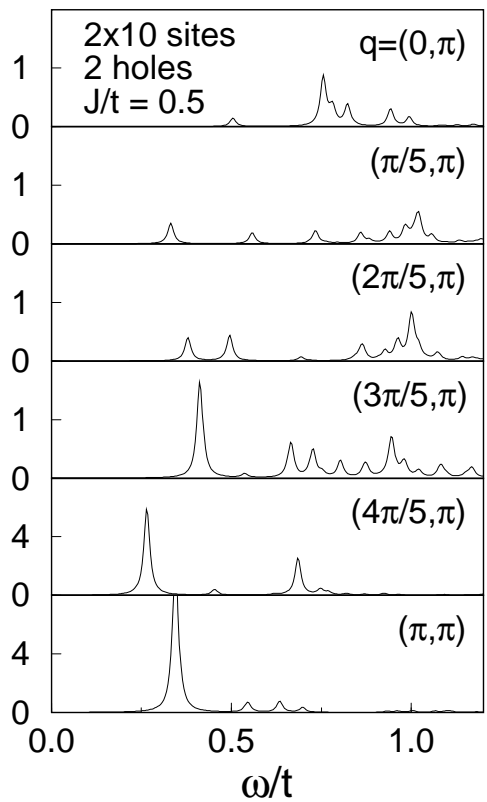
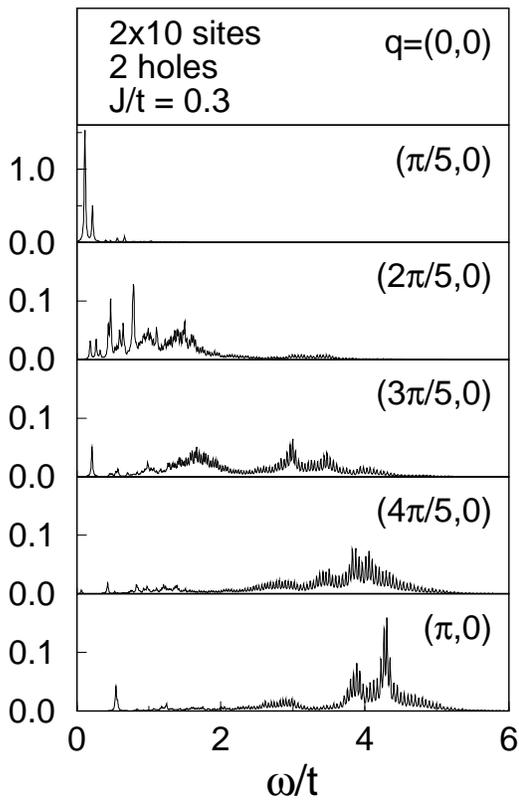
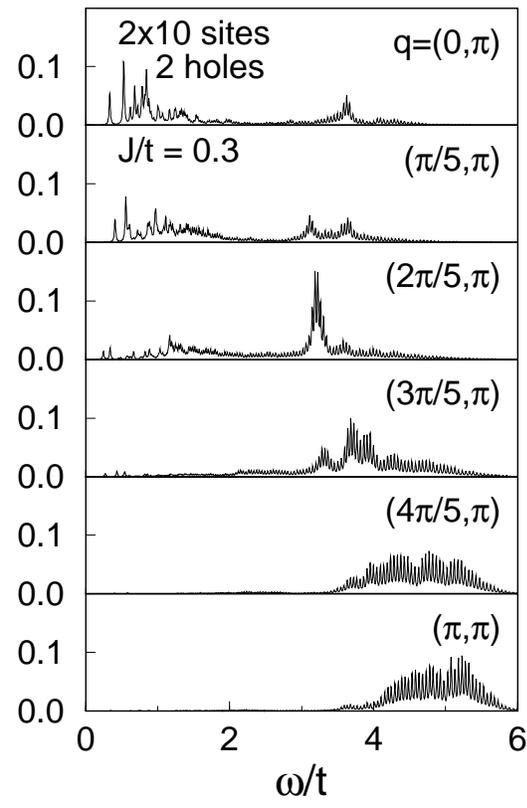
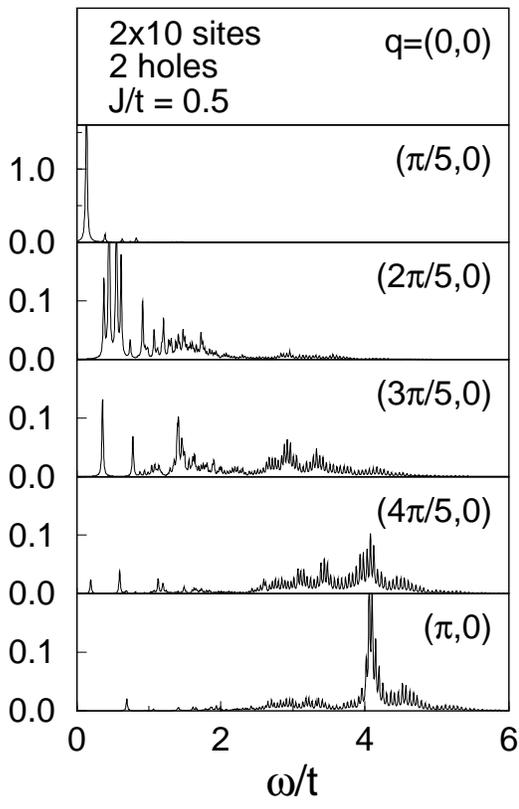
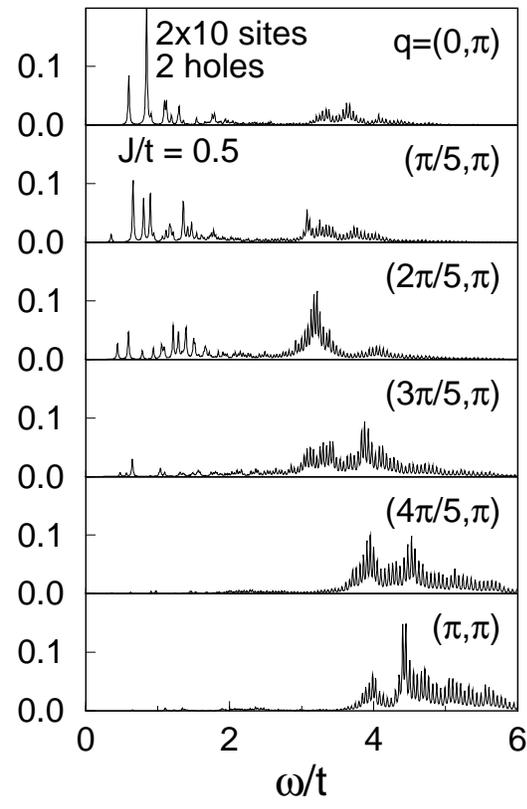
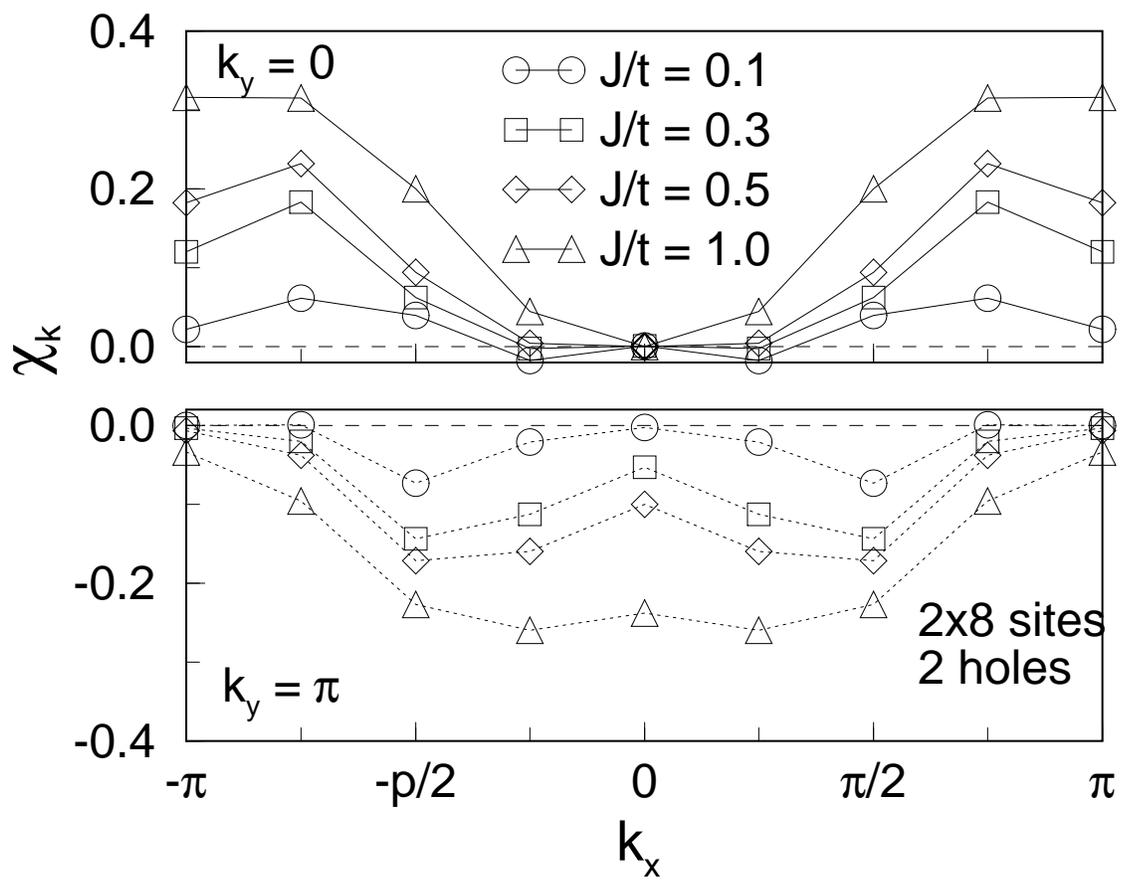
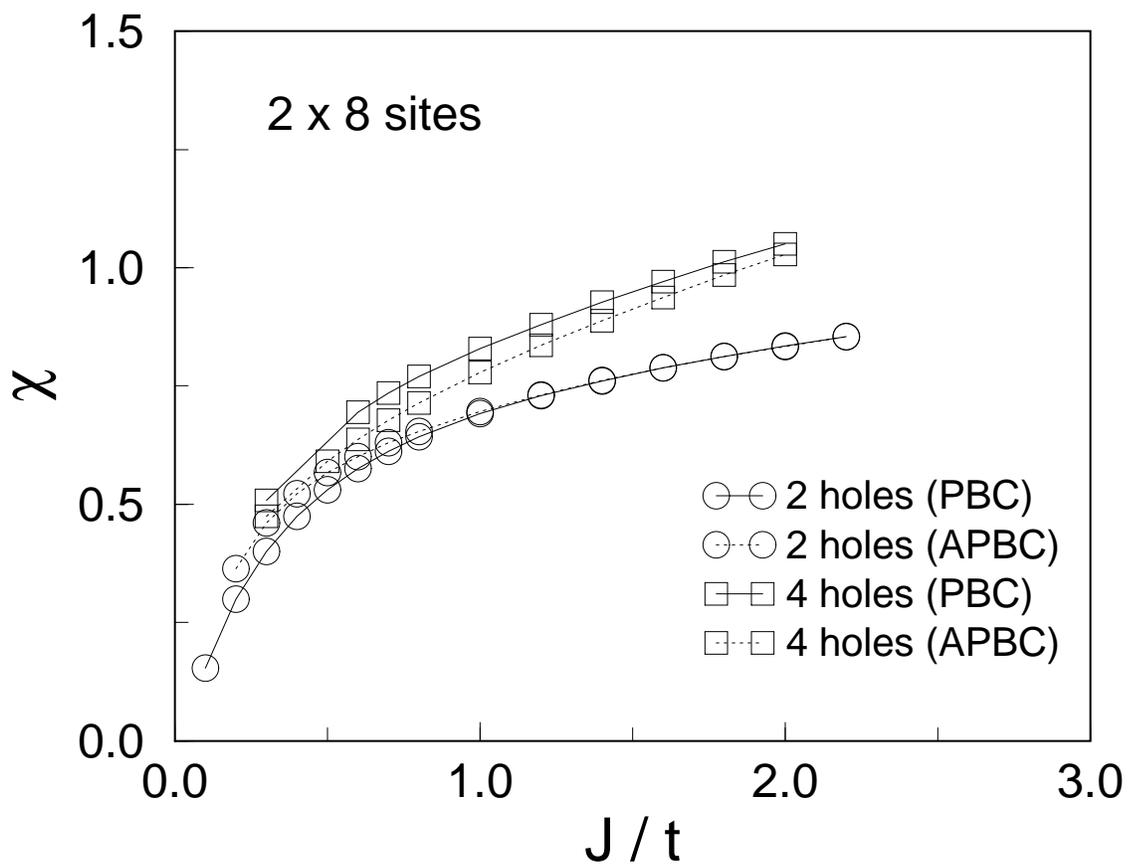
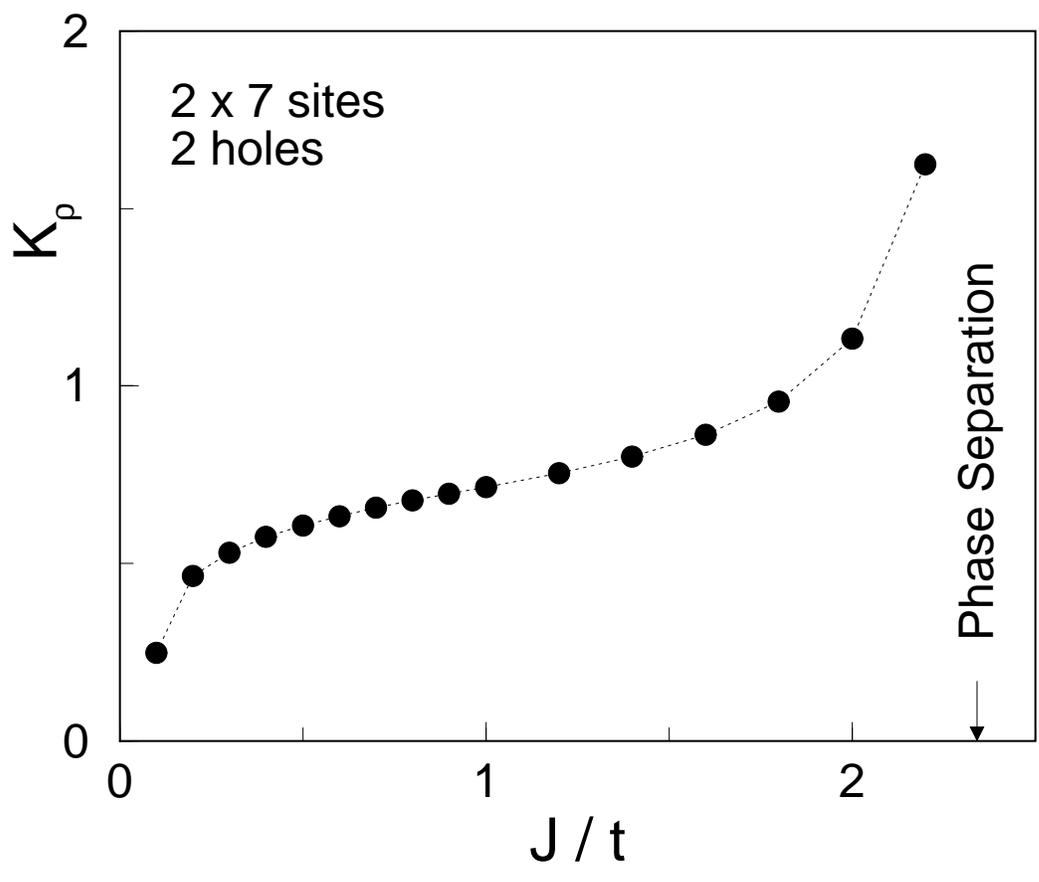
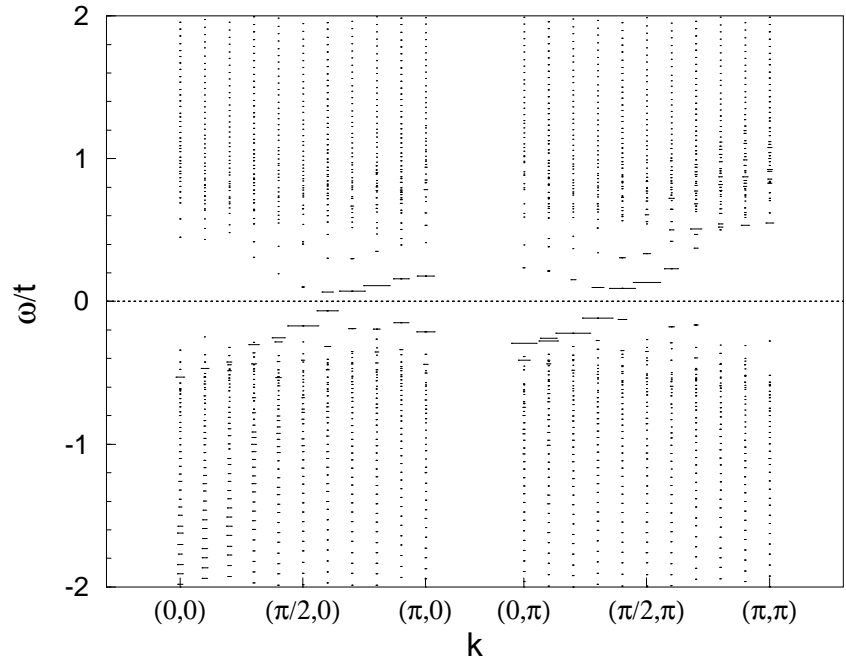
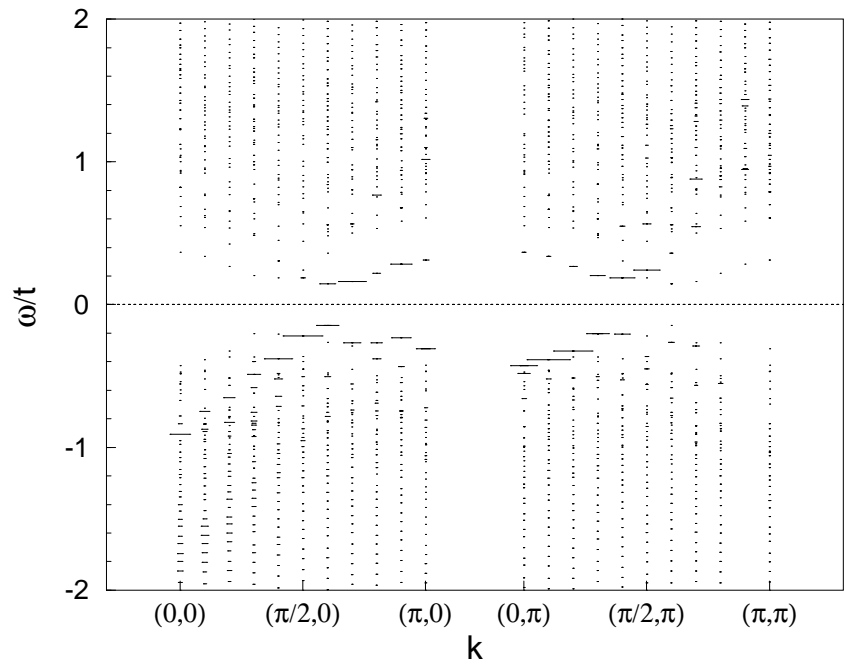
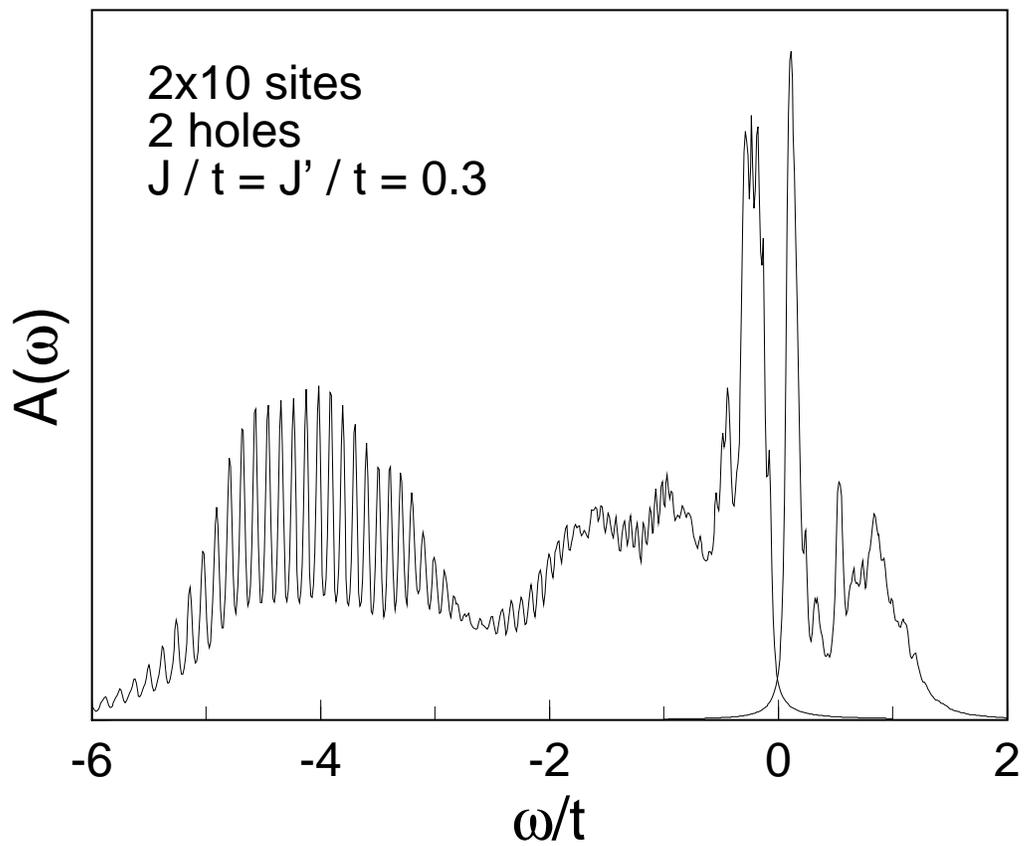
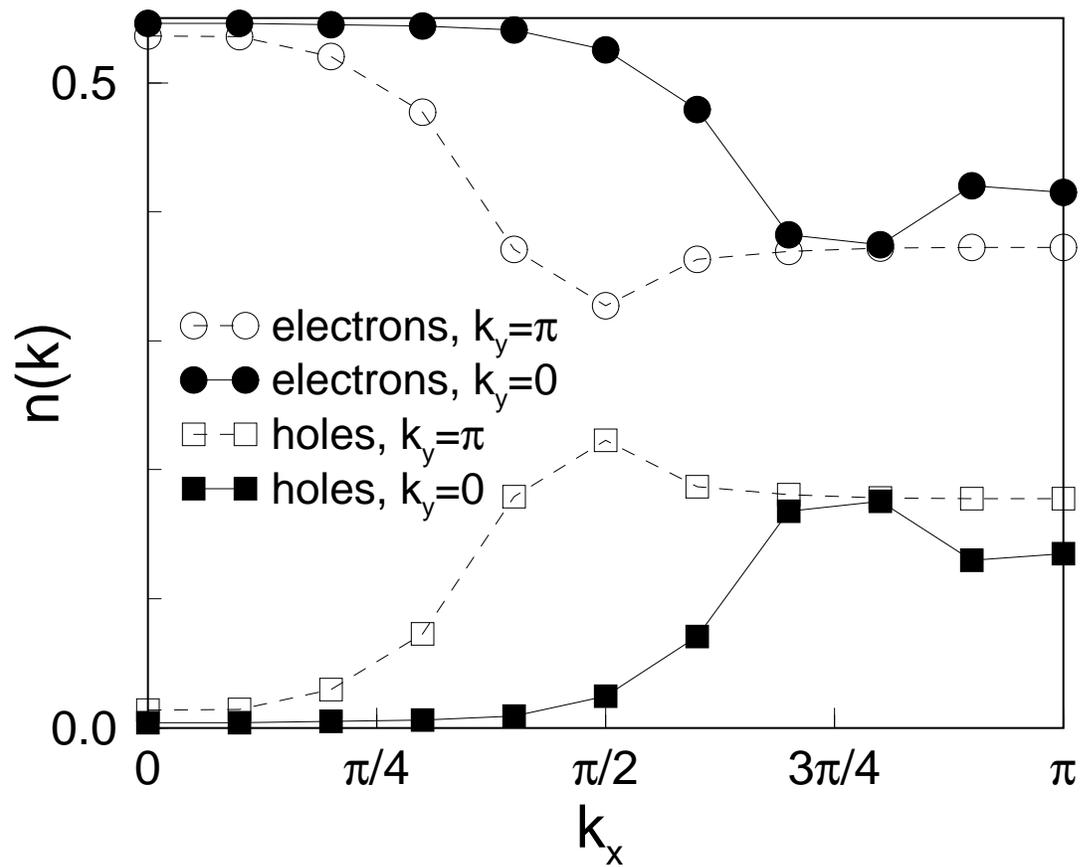
\end{document}